\documentclass{article}

\usepackage{jcappub}

\usepackage[utf8]{inputenc} 
\usepackage[T1]{fontenc}    
\usepackage{hyperref}       
\usepackage{url}            
\usepackage{booktabs}       
\usepackage{amsfonts}       
\usepackage{nicefrac}       
\usepackage{microtype}      
\usepackage{lipsum}
\usepackage{comment}
\usepackage{amsmath,amsfonts,amsthm,bm}
\usepackage{accents}
\usepackage[normalem]{ulem}
\usepackage{siunitx}
\usepackage{tabularx}
\usepackage{booktabs}
\usepackage{xcolor}
\usepackage{textgreek}
\usepackage{subcaption}
\usepackage{upgreek}
\usepackage{threeparttable} 
\usepackage{graphicx}
\usepackage{graphics}

\DeclareSIUnit\year{y}
\DeclareSIUnit\ton{t}
\DeclareSIUnit\kev{keV}
\DeclareSIUnit[per-mode=symbol]\gevm{\GeV\per\clight\squared}
\DeclareSIUnit[per-mode=symbol]\tevm{\TeV\per\clight\squared}
\DeclareSIUnit\tonneyears{\ton\times\year}
\newcommand{\dru}{\si{(keV.t.y)^{-1}}}

\newlength\mytemplength
\newcommand\parboxc[3]{%
    \settowidth{\mytemplength}{#3}%
    \parbox[#1][#2]{\mytemplength}{\centering #3}%
}

\usepackage{lineno}
    
\usepackage{currfile} 

\begin{document}

\title{Projected WIMP Sensitivity of the XENONnT Dark Matter Experiment}

\author{The XENON collaboration: }
\author[a]{E.~Aprile,}
\author[b]{J.~Aalbers,}
\author[c]{F.~Agostini,}
\author[d]{M.~Alfonsi,}
\author[e]{L.~Althueser,}
\author[f]{F.~D.~Amaro,}
\author[b]{V.~C.~Antochi,}
\author[g]{E.~Angelino,}
\author[h]{J.~R.~Angevaare,}
% ...
\affiliation[a]{Physics Department, Columbia University, New York, NY 10027, USA}
\affiliation[b]{Oskar Klein Centre, Department of Physics, Stockholm University, AlbaNova, Stockholm SE-10691, Sweden}
\affiliation[c]{Department of Physics and Astronomy, University of Bologna and INFN-Bologna, 40126 Bologna, Italy}
\affiliation[d]{Institut f\"ur Physik \& Exzellenzcluster PRISMA, Johannes Gutenberg-Universit\"at Mainz, 55099 Mainz, Germany}
\affiliation[e]{Institut f\"ur Kernphysik, Westf\"alische Wilhelms-Universit\"at M\"unster, 48149 M\"unster, Germany}
\affiliation[f]{LIBPhys, Department of Physics, University of Coimbra, 3004-516 Coimbra, Portugal}
\affiliation[g]{INAF-Astrophysical Observatory of Torino, Department of Physics, University  of  Torino and  INFN-Torino,  10125  Torino,  Italy}
\affiliation[h]{Nikhef and the University of Amsterdam, Science Park, 1098XG Amsterdam, Netherlands}
% ...

\emailAdd{pietro.digangi@bo.infn.it}
\emailAdd{jpienaar@uchicago.edu}
\emailAdd{diego.ramirez@physik.uni-freiburg.de}
\emailAdd{xenon@lngs.infn.it}

\keywords{Dark matter experiments, dark matter simulations}

\abstract{XENONnT is a dark matter direct detection experiment, utilizing 5.9\,t of instrumented liquid xenon, located at the INFN Laboratori Nazionali del Gran Sasso. In this work, we predict the experimental background and project the sensitivity of XENONnT to the detection of weakly interacting massive particles (WIMPs). The expected average differential background rate in the energy region of interest, corresponding to (1,\,13)\,keV and (4,\,50)\,keV for electronic and nuclear recoils, amounts to $12.3 \pm 0.6$\,\dru~and $(2.2\pm 0.5)\times 10^{-3}$\,\dru, respectively, in a 4\,t fiducial mass.  We compute unified confidence intervals using the profile construction method, in order to ensure proper coverage. With the exposure goal of 20\,\si{t.y}, the expected sensitivity to spin-independent WIMP-nucleon interactions reaches a cross-section of $1.4\times10^{-48}\,\mathrm{cm}^2$ for a 50\,GeV/c$^2$ mass WIMP at 90\% confidence level, more than one order of magnitude beyond the current best limit, set by XENON1T. In addition, we show that for a 50\,GeV/c$^2$ WIMP with cross-sections above $2.6\times10^{-48}\,\mathrm{cm}^2$ ($5.0\times10^{-48}\,\mathrm{cm}^2$) the median XENONnT discovery significance exceeds $3\sigma$ ($5\sigma$). The expected sensitivity to the spin-dependent WIMP coupling to neutrons (protons) reaches $2.2\times10^{-43}\,\mathrm{cm}^2$ ($6.0\times10^{-42}\,\mathrm{cm}^2$).}

\maketitle
\flushbottom

\section{Introduction}\label{Intro}

Astrophysical observations indicate that a significant fraction of the energy content of the Universe is composed of cold dark matter~\cite{Aghanim:2018eyx}. 
The most promising candidates for a particle explanation of dark matter are weakly interacting massive particles (WIMPs)~\cite{Roszkowski:2017nbc}. Over the past three decades a large number of search campaigns in underground laboratories have been conducted using a variety of techniques to detect these particles, which are expected to interact very rarely. 

Dual-phase liquid-gas xenon time projection chambers (TPCs)~\cite{Aprile:2018dbl, Cui:2017nnn, Akerib:2016vxi} are the world-leading detector technology for direct detection of WIMPs~\cite{Schumann_2019}. Liquid xenon (LXe) makes an ideal WIMP target due to its high stopping power for gamma and beta radiation, providing self-shielding from external backgrounds. Moreover, the absence of long-lived isotopes detrimental to WIMP searches minimizes the internal backgrounds.
The large atomic mass (A\,$\approx$\,131) enhances the expected rate of coherent scattering by WIMPs off the xenon nuclei.

The largest and most sensitive LXe detector to date is the XENON1T experiment~\cite{Aprile:2017aty}, which was situated at an average depth of 3600\,m water equivalent at the INFN Laboratori Nazionali del Gran Sasso.  XENON1T set the world's strongest limits on the spin-independent (SI) WIMP-nucleon coupling for almost all WIMP masses $>100$\,MeV/c$^{2}$~\cite{Aprile:2018dbl, Aprile:2019xxb, Aprile:2019jmx}. New parameter space has also been excluded for the spin-dependent (SD) WIMP-neutron interactions in the range of WIMP masses $>3$\,GeV/c$^{2}$. With the upgrade to the XENONnT experiment, we increase the instrumented LXe mass by a factor 3 and utilize most of the infrastructure already developed for XENON1T. The ultra-low XENON1T background level~\cite{aprile2020lowER}, the lowest ever achieved in dark matter LXe experiments, will be further reduced in XENONnT by the improved purity of the xenon inventory and the addition of a new neutron veto (NV). The NV will enable identification of otherwise irreducible neutron backgrounds in the target volume. XENONnT is expected to start taking science data in 2020.

In this work, we present the projected sensitivity of XENONnT to SI and SD WIMP-nucleon interactions based on a detailed simulation of the experiment. The detector description and working principle are presented in section~\ref{DetectorGeometry}, while the simulation of the detector response to particle interactions is discussed in section~\ref{sec:detector-response-model}. The relevant background contributions are assessed in section~\ref{sec:backgrounds}. The sensitivity of XENONnT in the search for both SI and SD WIMP-nucleon couplings is presented in section~\ref{sec:sensitivity}, for an exposure goal of \SI{20}{\ton\year}.
\section{The XENONnT experiment}\label{DetectorGeometry}

XENONnT consists of three nested detectors. It is enclosed by a cylindrical stainless steel (SS) tank, 10.2 m-high with a diameter of 9.6 m, filled with Gd-loaded water. The tank is instrumented with photomultiplier tubes (PMTs) and acts as a water Cherenkov muon veto. A second detector, the neutron veto, is contained within and optically separated from the muon veto volume. Finally, the LXe TPC is located at the center of the neutron veto system.

\subsection*{Time Projection Chamber} \label{DetectorModel}

Scintillation photons (through de-excitation of $\rm{Xe}_2$ dimers) and free electrons (via atomic ionization) are produced following energy depositions in LXe~\cite{RevModPhys.82.2053}. The prompt scintillation signal (S1) is observed by two arrays of photomultiplier tubes (PMTs) at the top and bottom of the TPC. The ionization electrons are drifted towards the top by means of an electric drift field applied across the active target. A strong electric field extracts the drifted electrons into the gaseous xenon (GXe) layer, present between the LXe volume and the top PMT array, where they produce proportional scintillation light (S2). The three-dimensional position of the interaction vertex is inferred from the localized hit pattern of the S2 light on the top PMT array (x-y position) and from the time difference between the S2 and the S1 due to the drift time of ionization electrons (depth, z position). The energy released in the detector is reconstructed combining the S1 and S2 signals~\cite{aprile2020energy}.

We distinguish two types of events in the LXe target: electronic recoils (ERs), produced by particles scattering off atomic electrons, and nuclear recoils (NRs) from scatters off xenon nuclei. The expected WIMP signature is a single low-energy (<\,50\,keV) NR. Due to the differing relative scintillation and ionization yields of ERs and NRs, a larger S2/S1 ratio is observed for ERs for the same energy deposition. Typically, ER discrimination powers greater than 99.5\% with 50\% NR acceptances have been achieved with the current generation of xenon dual-phase TPCs~\cite{Akerib:2013tjd, Cui:2017nnn,Aprile:2019dme}.

\begin{figure}[t]
\centering 
\includegraphics[width=.45\textwidth]{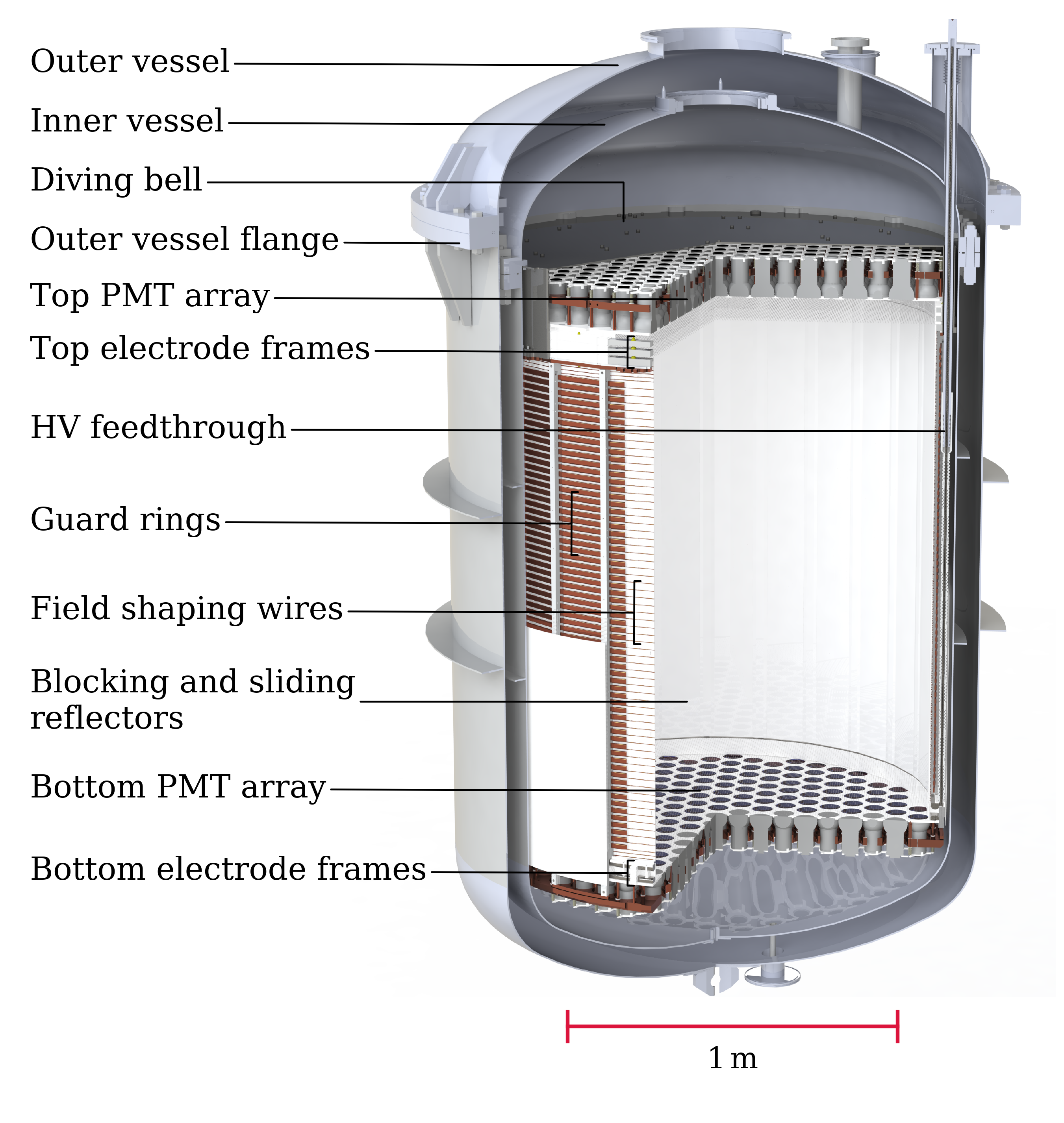}
\hfill
\includegraphics[width=.54\textwidth]{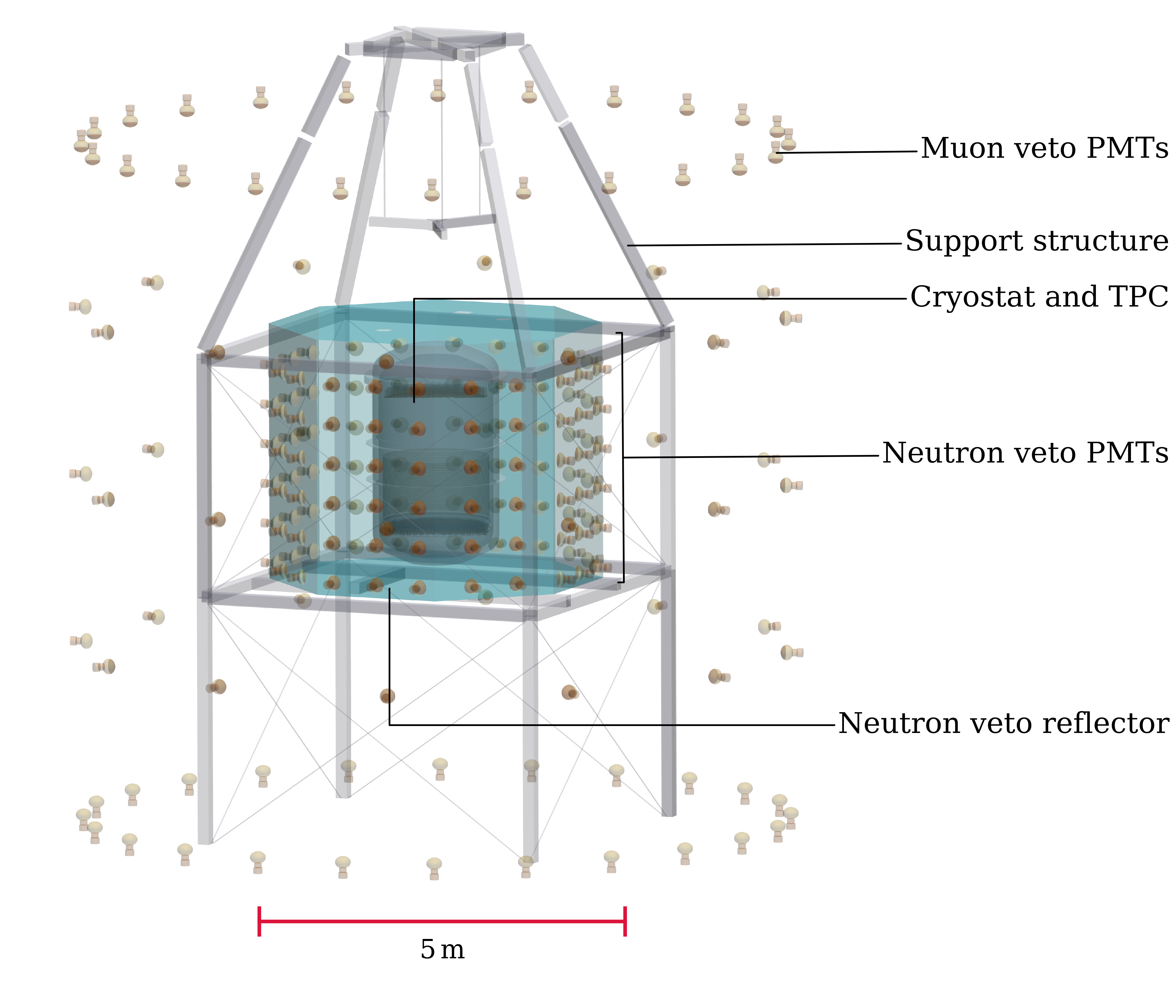}
\caption{\label{fig:cad_model} 
(Left) CAD rendering of the XENONnT cryostat and TPC. The TPC has a diameter of 1.3\,m and is 1.5\,m-tall. (Right) Geant4 rendering of the three nested detectors, including muon and neutron veto. The water tank walls, which support the muon veto PMTs, the neutron veto support structure, and other components (e.g. calibration systems) are omitted for clarity. Reflector panels, which optically separate the neutron and muon vetos, are shown as transparent turquoise surfaces. The neutron veto PMT windows face the neutron veto region through openings in the panels.}
\end{figure}

A rendering of the XENONnT TPC inside the double-walled cryostat is shown in figure~\ref{fig:cad_model} (left). The active region of the TPC contains 5.9\,t of LXe, enclosed by 24 polytetrafluoroethylene (PTFE) reflector panels. The polygonal shape formed by the reflector panels has an apothem of 664\,mm. The corners of the active region are covered by an additional set of 24 blocking panels, which overlap with the reflector panels to optically isolate the active region from the surrounding LXe. The top (bottom) PMT array consists of 253 (241) PMTs, arranged in a compact hexagonal structure to maximize light collection efficiency (LCE). The PMTs are 3'' Hamamatsu R11410-21, chosen for their low radioactivity~\cite{Aprile:2015lha} and high quantum efficiency (QE\,$\approx$\,34\% on average at room temperature) at the xenon scintillation wavelength of \SI{175}{\nano\meter}~\cite{FUJII2015293}. PMT characterization and performance are discussed in refs.~\cite{Barrow:2016doe, Baudis:2013xva}.

As in previous XENON detectors, the liquid level in the TPC is controlled by means of a SS diving bell. The drift field is generated by means of a gate electrode slightly below the liquid-gas interface and a cathode. The active region is demarcated by these two electrodes, which are separated by 1485\,mm at operating temperature. An anode electrode is placed in the GXe 8\,mm above the gate. In addition to the anode, gate and cathode electrodes, the TPC has two screening electrodes. These are positioned directly below (above) the top (bottom) PMT array to screen the PMTs from the field produced by the anode (cathode). The electrodes consist of parallel SS wires, which are \SI{216}{\micro\meter}-thick with the exception of the cathode (\SI{304}{\micro\meter}), stretched onto SS rings. The top electrodes (top screening, anode, gate) have a pitch of 5\,mm, while the bottom electrodes (cathode, bottom screening) have a pitch of 7.5\,mm. The gate and anode have two and four additional \SI{304}{\micro\meter}-thick wires, respectively, running perpendicularly to all other wires. These perpendicular wires are added to counteract deformation of the electrode plane under electrostatic forces.

Uniformity of the drift field is achieved by two concentric sets of OFHC copper field shaping rings, vertically interleaved and with a 15\,mm radial separation. The inner set consists of 71 field shaping wires of 2\,mm diameter and touches the outer side of the PTFE panels of the TPC. The outer set is made of 64 rings, which are 15\,mm-tall and 5\,mm-thick.

The full 8.4\,t LXe inventory is contained in a double-walled  vacuum-isolated cryostat, consisting of an inner and outer vessel, each with a domed upper section penetrated by several access ports. Two double-walled vacuum-insulated pipes run from the largest access ports to the cryogenics and purification systems in the adjacent service building. These pipes also house the signal and HV cables of the PMTs, sensors, and electrodes with the exception of the cathode. Two smaller ports accommodate a motion feed-through to level the TPC and a feedthrough for the cathode high voltage supply. The lower section of each vessel is a cylinder capped with a dome at the bottom. A single port at the bottom allows for LXe purification and fast LXe recovery. The vessel walls are made of 5\,mm-thick low-radioactivity SS, while the upper and lower sections of each vessel are mated by 45\,mm-thick SS flanges. 

Electronegative impurities in the LXe target can trap ionization electrons, reducing the observed amplitude of S2 signals. In addition to the existing GXe purification system with increased purification flow with respect to XENON1T, LXe is constantly circulated through a novel liquid purification system. Radioactive contaminants in the LXe, such as krypton and radon, will contribute to the background. Krypton is removed by means of cryogenic distillation through a dedicated column already used for XENON1T~\cite{Aprile:2016xhi}. A newly developed radon distillation column will further suppress radon backgrounds, based on the principle demonstrated in ref.~\cite{xe1t-rnpaper}.

\subsection*{Neutron Veto} \label{sec:neutron_veto}

The XENONnT neutron veto (NV) will reduce the radiogenic neutron background by tagging events where the interaction in the TPC is coincident with a neutron detected in the NV. A total of 120 Hamamatsu R5912-100-10 8'' high-QE (40\% on average at \SI{350}{\nano\meter}) PMTs with low-radioactivity windows are placed along reflective panels around the cryostat. The NV lateral panels form an octagonal enclosure with an apothem of 2\,m and a height of 3\,m. The PMTs are distributed among 6 equally-spaced rows with only the PMT window protruding into the NV region. The bulk of the PMT body remains behind the reflective panels in order to minimize the radioactive background inside the NV. Octagonal reflective end caps enclose the system at the top and bottom. All the reflective surfaces are made of 1.5\,mm-thick expanded PTFE (ePTFE), for which we measured a reflectivity to Cherenkov light greater than 99\% for wavelengths above 280\,nm. A rendering of the NV is shown in figure~\ref{fig:cad_model} (right).

Neutrons that scatter in the TPC volume can easily pass through the cryostat and escape further detection in LXe. In order to enhance the neutron detection probability via neutron capture, the water within the muon veto tank is loaded with gadolinium sulphate octahydrate ($\rm{Gd}_2(\rm{SO}_4)_3 \cdot 8(\rm{H}_2\rm{O})$), providing a 0.2\% Gd relative mass concentration. Neutrons that leave the TPC volume will be moderated by the water around the cryostat, typically travelling less than 20\,cm before being thermalized and captured by Gd (H) with a probability of 91\% (9\%). Following the neutron capture by Gd, a gamma-ray cascade with total energy of about 8\,MeV is generated. In the case of capture on H, a single 2.2\,MeV gamma is emitted. The energy deposited by the gammas in the water, mainly through Compton scattering, is converted into electrons and ultimately into Cherenkov photons.

The feasibility of this neutron detection scheme for use in Super-Kamiokande has been demonstrated by the EGADS project, which showed that an absorption length compatible with that of pure water can be maintained in Gd-loaded water~\cite{Ikeda:2019pcm}. Accordingly, in this work we assume the Super-Kamiokande absorption length of $\mathcal{O}$(100\,m)~\cite{Abe:2013gga}. Photons in the NV volume may thus reflect multiple times before hitting a photosensor. For this reason, the LCE is mostly independent of the geometrical arrangement of the PMTs or their distance from the cryostat, but relies mainly on the total photosensitive area, and thus the number of PMTs. To maximize the LCE, the outer vessel of the cryostat is clad with ePTFE as well.

\subsection*{Muon Veto} \label{sec:muon_veto}
The muon veto system, containing $\sim$\,700\,t of Gd-loaded water, is inherited from XENON1T, where the SS tank was filled with pure deionized water. It is instrumented with 84 Hamamatsu R5912ASSY 8'' PMTs and operated as an active Cherenkov muon veto able to tag incoming muons and hadronic showers produced by muon-induced spallation reactions in the cavern rock. Additionally, the water provides effective shielding against environmental gamma and neutron radiation. Detailed information about the muon veto can be found in ref.~\cite{Aprile:2014zvw}.
\section{Simulation framework}\label{sec:detector-response-model}

The XENONnT detector is simulated with the Geant4 toolkit~\cite{Agostinelli:2003geant4, ALLISON2016186}. The Monte Carlo (MC) framework is built upon the XENON1T simulation package~\cite{Aprile:2015uzo}. Interactions in the detector are studied by simulating particle generation and propagation through the detector volumes. Energy depositions in the TPC are converted into S1 and S2 signals in order to evaluate the expected background and signal distributions in the observable space. This conversion is based on the model of light and charge emission following an interaction in LXe, convoluted with detector effects related to signal collection efficiency and reconstruction.

\subsection{Particle generation and propagation} \label{SimulationSoftware}

Geant4 version 10.3-patch03 is used for XENONnT simulations. Radioactive decays are simulated via the \textit{G4RadioactiveDecay} process and, if the daughter of the nuclear decay is an isomer, prompt de-excitation is handled by the \textit{G4PhotonEvaporation} process, where the relevant parameters (half-lives, nuclear level structure, decay branching ratios, and energy) are taken from the Evaluated Nuclear Structure Data Files (ENSDF)~\cite{ensdf}. The \textit{Livermore} physics list is used for high precision tracking of gamma and electron interactions. The \textit{QGSP\texttt{\_}BERT\texttt{\_}HP} hadronic physics list provides high-precision data-driven models for the scattering and capture processes of neutrons at low energies ($<20$\,MeV), using the G4NDL4.5 neutron library with thermal cross-sections~\cite{Mendoza:2018g4ndl}. In addition, the inaccurate default Geant4 modeling of the gamma emission after neutron capture by Gd is changed to the data-driven description provided in refs.~\cite{Tanaka:2019hti, Hagiwara:2018kmr}, correcting for both energy conservation during de-excitation and gamma multiplicity of the cascade. The default models for neutron capture by other nuclei are not modified. The generation of Cherenkov photons after neutron-induced signals is included in the event-by-event simulations of neutron signals, accounting for the neutron generation and recoil in the active volume up to the detection of these photons by the NV PMTs.

\subsection{Liquid xenon signal response}\label{sec:lxe-effects}

The energy released in LXe by an incident particle via ionization and diatomic de-excitation yield detectable quanta: free electrons and photons, respectively. Their emission is characterized following the NEST (Noble Element Simulation Technique) parameterization~\cite{Lenardo:2014cva}, taking into account fluctuations in the scintillation and ionization processes, electron-ion recombination, and drift field dependence. Specifically, we adopt the detailed LXe emission model fitted to XENON1T high-statistics calibration data, described in ref.~\cite{Aprile:2019dme}. The photon and electron yields below 1\,keV for ER and 3.5\,keV for NR are extrapolated based on measurements at higher energies, where zero emission below 1\,keV NR is assumed.

\subsection{Detector effects}\label{sec:detector-effects}

\begin{table}[t]
\centering
\begin{tabular}{lr}
\toprule
TPC parameters & Value\\
\midrule
\multicolumn{2}{l}{\textbf{Optical parameters}}\\
PTFE-LXe (GXe) reflectivity & 0.99 (0.99)\\
LXe absorption length [m] & 50\\
LXe Rayleigh scattering length [cm] & 30 \\
LXe (GXe) refractive index & 1.63 (1) \\
PMT quartz window refractive index & 1.59\\
Electrodes optical transparency \\
\hspace{3mm} Top screen & 0.957 \\
\hspace{3mm} Anode  & 0.956 \\
\hspace{3mm} Gate  & 0.956 \\
\hspace{3mm} Cathode & 0.960 \\
\hspace{3mm} Bottom screen & 0.971 \\
\midrule
\multicolumn{2}{l}{\textbf{Signal generation}}\\
PMT quantum efficiency (QE) & 0.34\\
PMT collection efficiency~\cite{Lung:2012pi} & 0.90\\
Double photoelectron (PE) probability~\cite{Faham:2015kqa,Paredes:2018hxp} & 0.219 \\
Photon detection probability (g\textsubscript{1}) [PE/ph] & 0.169 \\
Electron extraction efficiency & 0.96\\
Effective charge gain (g\textsubscript{2b}) [PE/e\textsuperscript{-}] & 14.3 \\
S1 PMT coincidence level  & 3 \\
\midrule
\multicolumn{2}{l}{\textbf{Detector conditions}}\\
Drift field [V/cm] & 200 \\
Electron lifetime [\si{\micro\second}] & 1000 \\
\bottomrule
\end{tabular}
\caption{\label{tab:det-parameters}TPC parameters used in the XENONnT detector response model. The average PMT QE at room temperature and the wavelength-dependent optical parameters are given for the xenon scintillation wavelength of \SI{175}{\nano\meter}. Although the purity of the xenon target is expected to be higher, thanks to the upgraded purification system, the LXe absorption length is conservatively taken from XENON1T.}
\end{table}

\begin{figure}[t]
    \centering
    \includegraphics[width=\textwidth]{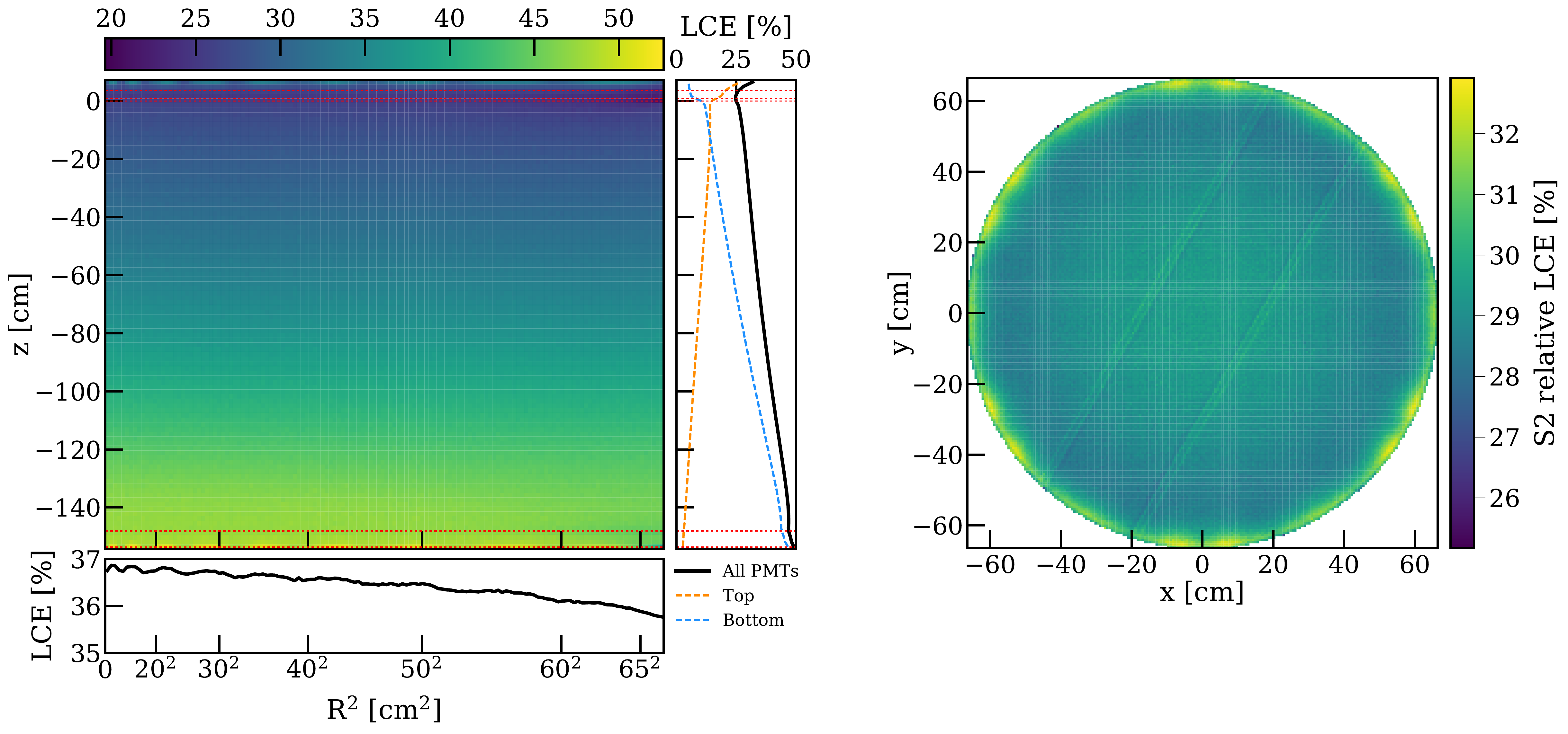}
    \caption{\label{fig:lce} (Left) Light collection efficiency (LCE) map of the XENONnT detector. The red dotted lines correspond to the position of the electrodes within the TPC, with the gate situated at z\,=\,0. The bottom and right panels show the LCE variation along the R and z coordinates, respectively. (Right) Fraction of total detected S2 light seen by the bottom PMT array. Simulations are performed by generating S2 photons in the GXe region below the anode. Line features visible are due to the perpendicular wires on the gate and anode. Effects from other wires are smeared out as the wire pitch is smaller than the binning of the histogram.}
\end{figure}

S1 and S2 signals are simulated accounting for the expected detector conditions. The main detector parameters of the XENONnT TPC are listed in table~\ref{tab:det-parameters}. The propagation of the prompt scintillation photons (S1) in the TPC is simulated with Geant4 in order to estimate the light collection efficiency (LCE) at the photocathode of the PMTs. The optical simulation framework is detailed in ref.~\cite{Aprile:2015uzo}, together with a description of the optical properties of the detector materials. The assumptions for LXe, GXe and the PMT window, listed in table~\ref{tab:det-parameters}, are the same as for XENON1T. The five XENONnT electrodes consist of parallel wires, which are implemented in detail in the simulated detector geometry. The values of their optical transparency, defined as the fraction of non-opaque surface area, are included in table~\ref{tab:det-parameters} for reference.

The XENONnT LCE map is shown in figure~\ref{fig:lce} (left). The average LCE in the active region of the TPC is 36\%, ranging from a maximum of $\sim$\,50\% just above the cathode to $\sim$\,25\% in the region right below the GXe, due to internal reflection at the liquid-gas interface. The relative variation over the TPC radius is within 3\%. The overall LCE is slightly higher than in XENON1T, despite the higher rate of absorbed photons in the larger LXe volume, a result of the more compact top PMT array and more transparent electrodes. The corrected S1 signal (cS1) accounts for the photon detection probability (g\textsubscript{1}), which is the product of the position-dependent LCE, the QE of the PMTs at \SI{175}{\nano\meter} and LXe temperature~\cite{Lyashenko_2014}, and the collection efficiency at the first PMT dynode. In our simulated detector model, we account for the relative differences in response at the per-PMT level due to the angle of incidence of a photon on the photocathode, as well as the spatially non-uniform response across the photocathode~\cite{Lung:2012pi}.

Electrons generated at the interaction point are drifted to the liquid-gas interface by the design drift field of 200\,V/cm. The signal amplitude at the interface is corrected for the probability of electron loss to electronegative impurities, parameterized by the target electron lifetime of 1000\,\si{\micro\second}. The corrected S2 signal (cS2) additionally accounts for the electron extraction efficiency into the GXe, the gas gain (number of photoelectrons per extracted electron), and the xy-dependent S2 LCE. The two-dimensional S2 LCE map of the bottom PMT array, shown in figure~\ref{fig:lce} (right), is obtained by estimating the detection efficiency for VUV photons generated in the S2 region right below the anode. In agreement with the method followed for the XENON1T WIMP search results~\cite{Aprile:2019bbb}, we use the corrected S2 detected by the bottom PMT array (cS2\textsubscript{b}) for energy reconstruction, given the more homogeneous LCE. Therefore, the average effective charge gain (g\textsubscript{2b}) of 14.3\,PE/e\textsuperscript{-} corresponds to the bottom array-only signal.

Resolving multiple interactions depends on the separation efficiency of the S2 signals. Simulated multiple recoils are clustered by applying a three-dimensional resolution map which depends on the S2 size and the spatial separation along the vertical z-axis.  We conservatively assume no separation power from the relative distance in the (x, y) plane. The minimum z separation ranges from approximately 3\,mm, for small signals produced in the top of the TPC, to 20\,mm for large S2 amplitudes near the cathode. The parameterization of the multiple recoil resolution is based on XENON1T raw waveform simulations, which were validated with $^{241}$AmBe neutron calibration data~\cite{Aprile:2019bbb}.
\section{Backgrounds} \label{sec:backgrounds}
Backgrounds from sources producing either ERs or NRs are estimated via MC simulations of the recoiling particles in the LXe target, accounting for the multiple recoil resolution to assess the multiplicity of the events in the TPC. The WIMP signal is expected to be a single NR interaction in the detector volume, thus our event selection is restricted to single scatters.
Although ERs can be efficiently discriminated from NRs based on the S2/S1 ratio, statistical leakage of the ER population can still produce events indistinguishable from WIMPs. Thus, a detailed understanding of both NR and ER background sources is required. 
We estimate the rate of NR backgrounds in the (4,\,50)\,keV energy range of interest (energy ROI), which corresponds to (1,\,13)\,keV for ERs in the S1 signal space.

Radioactive isotopes dissolved in the xenon itself, such as $^{222}$Rn and $^{85}$Kr, are sources of intrinsic ER background uniformly distributed in the active volume and thus unmitigated by the LXe self-shielding power. The dominant background in XENON1T was caused by the beta-emitter $^{214}$Pb, a product of the $^{222}$Rn decay chain. In XENONnT, the $^{222}$Rn level will be reduced by meticulous selection of low radon-emanating materials, detector design, smaller surface-area-to-volume ratio, and a dedicated online radon distillation column, a concept demonstrated in a dedicated experiment~\cite{Bruenner:2016ziq}, and tested in XENON100~\cite{Aprile:2017kop} and XENON1T~\cite{xe1t-rnpaper, murra2018}.

Irreducible backgrounds arise from interactions induced by neutrinos of solar, atmospheric, or supernova origin. These are spatially uniform due to the small cross-sections involved. Elastic scattering off xenon electrons contributes to the ER background, while coherent elastic neutrino-nucleus scattering (CE$\nu$NS) is responsible for an NR background.

Traces of radioactive isotopes in the detector components close to the active LXe volume can lead to both ER and NR backgrounds. Restricting searches to an inner fiducial volume (FV), effectively shielded by the outer LXe layer, reduces these external backgrounds as they mostly affect the outer TPC volume. We select a cylindrical FV containing a LXe mass of 4\,t, whose radius and height are optimized based on the spatial distribution of the materials background. ``Surface'' events at the PTFE walls can also constitute a background, as observed in XENON1T~\cite{Aprile:2018dbl}. We conservatively choose the FV shape, with bounds approximately 6\,cm away from the TPC walls, in order to suppress this contribution.

\subsection{Radioassay of detector components} \label{radioassay}

All materials used to build XENONnT were selected in a thorough radioassay program~\cite{nT_radioassay}.
High-purity germanium detectors~\cite{gator, gempi, gemse} measured the specific activities of the relevant gamma-ray emitters, including the primordial $^{40}$K isotope and $^{238}$U and $^{232}$Th chains. Gamma-ray spectroscopy allows us to detect a break of secular equilibrium, therefore the treatment of the early and late primordial chains is handled separately, as in refs.~\cite{Aprile:2013tov,Aprile:2015uzo}. Complementarily, high-resolution inductively coupled plasma-mass spectrometry (ICP-MS)~\cite{NISI2009828, Vacri:2013mfz} was used to accurately measure the amount of $^{238}$U and $^{232}$Th. The selection of components that are in contact with the LXe inventory is also based on radon emanation measurements~\cite{xe1t-rnpaper}, to ensure the lowest possible $^{222}$Rn contamination of the LXe target.

\begin{table}[tbp]
    \captionsetup{font=large}
    \resizebox{\textwidth}{!}{%
    \centering
    \begin{threeparttable}
    \begin{tabular}{lrcccccccc}
    \toprule
    Component & Mass & \multicolumn{8}{c}{Activity [mBq/kg]}\\\cline{3-10}
    & [kg] & \parboxc{c}{3ex}{$^{238}$U} & \parboxc{c}{3ex}{$^{235}$U} & \parboxc{c}{3ex}{$^{226}$Ra} & \parboxc{c}{3ex}{$^{232}$Th} & \parboxc{c}{3ex}{$^{228}$Th} & \parboxc{c}{3ex}{$^{60}$Co} & \parboxc{c}{3ex}{$^{40}$K} & \parboxc{c}{3ex}{$^{137}$Cs}\\
    \midrule
    Cryostat vessels & 1120 & 3.2\,(9) & 0.37\,(13) & 0.37\,(5) & 0.29\,(7) & 0.45\,(5) & 2.5\,(5) & 2.1\,(3) & <\,0.41\\
    Cryostat flanges & 730 & 1.4\,(4) & 0.06\,(2) & <\,4 & 0.21\,(6) & 4.5\,(6) & 14.1\,(9) & <\,5.6 & <\,1.5\\
    Bell and electrodes$^{ (1)}$ & 190 & 3.2\,(7) & 0.57\,(10) & 0.62\,(10) & 0.36\,(14) & 0.46\,(9) & 0.78\,(11) & 1.6\,(6) & <\,0.17\\
    PTFE$^{ (2)}$ & 128 & 0.12\,(5) & <\,0.06 & 0.10\,(2) & 0.11\,(5) & <\,0.06 & <\,0.053 & 2.4\,(3) & <\,0.038\\
    Copper$^{ (3)}$ & 355 & <\,0.69 & <\,0.28 & 0.033\,(5) & <\,0.027 & <\,0.023 & 0.11\,(2) & <\,0.29 & <\,0.016\\
    PMTs and bases$^{ (4)}$ & 98 & 53\,(15) & 2.2\,(7) & 4.6\,(10) & 3.5\,(12) & 4.2\,(8) & 7.1\,(9) & 73\,(18) & 0.9\,(3)\\
    \bottomrule
    \end{tabular}
    \caption{\label{tab:screening} 
    Radioactivity levels of the XENONnT detector components, with uncertainties in parenthesis.
    Upper limits are given at 90\% confidence level.
    The activities are averaged by mass over all the individually simulated sub-components. Omitted components, including those outside of the cryostat, induce less than 4\% of the total background rate from detector materials.
    $^{(1)}$SS diving bell and SS frames of the electrodes. 
    $^{(2)}$TPC pillars, blocking and sliding reflector panels, and PMT holders. 
    $^{(3)}$Support structure of the PMT arrays, support rings of the TPC, inner and outer field shaping rings.
    $^{(4)}$The total mass corresponds to 494 PMTs and PMT bases.}
    \end{threeparttable}}
\end{table}

The activity levels of the relevant components considered in the MC simulations are summarized in table~\ref{tab:screening}. In total, we simulate background contributions from 28 different detector components. The ER (NR) background contribution from each material is estimated by generating up to 10$^{9}$   (10$^{7}$) decays (neutrons) per isotope included in table~\ref{tab:screening}.  We conservatively assume the quoted 90\% confidence level (CL) upper limits as detection values. The majority of XENON1T systems outside the TPC have been reused for XENONnT, including major background sources such as the outer cryostat vessel. The contamination values for these components have been taken from our previous radioassay campaign~\cite{Aprile:2017ilq}. This also applies for 178 PMTs in the TPC, which are re-used from XENON1T~\cite{Aprile:2015lha}. As detailed in ref.~\cite{Aprile:2015uzo}, we follow two different strategies when determining the contribution of PMTs to the background: for ERs we simulate the contribution of the entire PMT, while for NRs, due to the material-dependent neutron yield, we estimate the contribution from the individual components of the PMTs separately: window (Quartz), body (Kovar; a cobalt-free Fe-Ni alloy), stem (Al$_{2}$O$_{3}$), SS parts, and bases (Cirlex; C$_{22}$H$_{10}$N$_{2}$O$_{5}$).  

The combined XENON1T and XENONnT screening campaign has been conducted over more than seven years. Consequently, the measured activities are rescaled to the values reported in table~\ref{tab:screening}, in order to account for the decay of all isotopes up to May 1, 2020. The predicted background rates from detector radioactivity correspond to the average over a five-year exposure from that date. Due to its 5.3\,y half-life, the rescaling mostly affects the event rate from $^{60}$Co, which is the largest contributor to the ER background from materials.

\subsection{Electronic recoil background}\label{sec:er-background}

\subsubsection*{Detector components}
\begin{figure}[tbp]
    \centering
    \includegraphics[width=\textwidth]{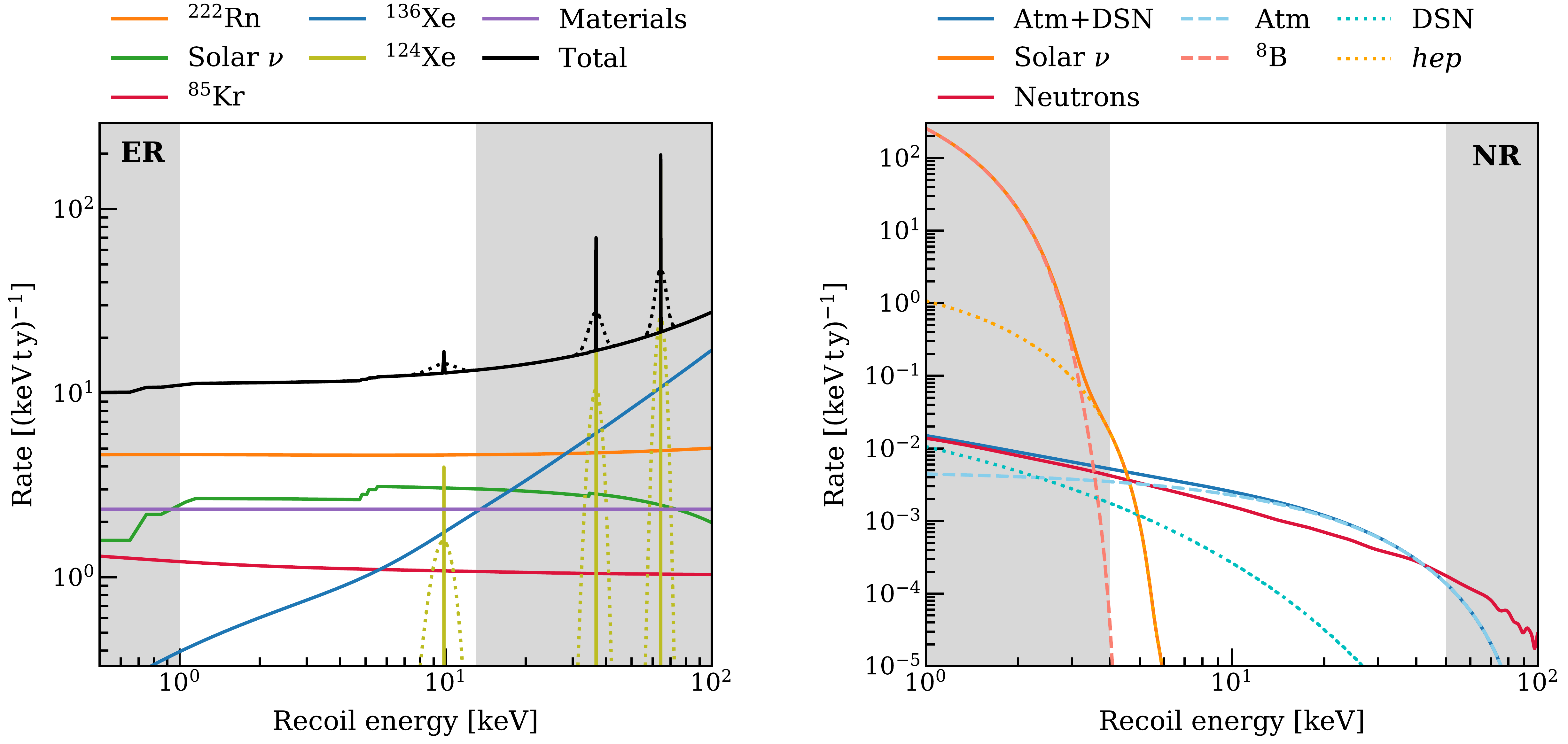}
    \caption{\label{fig:er-spectra} Energy spectra of the ER and NR backgrounds in the 4\,t fiducial volume of the XENONnT detector. Unshaded areas correspond to the WIMP search energy ROI: (1,\,13) and (4,\,50)\,keV for ERs and NRs, respectively. (Left) The largest ER background is due to the \SI{1}{\micro\becquerel/\kilogram} activity concentration of $^{222}$Rn (orange line).
    Additional backgrounds arise from solar neutrinos (green), double-beta decay of $^{136}$Xe (blue), $^{85}$Kr (red) due to a 0.1\,ppt concentration of \textsuperscript{nat}Kr, double-electron capture in $^{124}$Xe (olive), and detector components (purple). A Gaussian smearing of the $^{124}$Xe spectrum, based on the XENON1T energy resolution~\cite{XENON:2019dti}, is shown in dotted lines. The total ER spectrum (solid black line) is used to produce the final ER background model presented in section~\ref{sec:sensitivity}. (Right) The NR background contributions come from radiogenic neutrons (red) and CE$\nu$NS of solar neutrinos (orange), specifically $^8$B (dashed) and \emph{hep} (dotted), atmospheric (dashed blue) and diffuse supernova neutrinos (dotted blue). The neutron spectrum accounts for the NV tagging efficiency.}
\end{figure}

Gamma radiation produced by radioactivity in detector components can contribute to the low-energy background if it produces a single Compton scatter in the active LXe volume. External X-rays cannot reach the inner volume as their penetration depth in LXe is $\mathcal{O}$(\SI{10}{\micro\meter}). Radioactive decays are simulated from parent nuclides distributed uniformly within the respective detector components. The energy spectrum of the induced electronic recoils is shown in purple in figure~\ref{fig:er-spectra} (left). At low energies ($<200$\,keV) the Compton spectrum is almost flat and the differential rate in the reference 4\,t FV amounts to 2.1\,\dru. We assume a 10\% systematic uncertainty on the rate prediction based on material radioactivity measurements and the statistical uncertainty related to the number of simulated decays.

The two SS cryostat vessels account for 41\% of the total background from materials, while 51\% comes from the PMTs. Radioactivity from the SS bell constitutes an additional 6\%, while contributions from the electrodes, PTFE and copper components account for less than 3\%.

\subsubsection*{Radon}
Due to its 3.8\,d half-life, the emanation of $^{222}$Rn from detector materials results in a distribution of $^{222}$Rn and its decay products within the whole LXe volume. The most significant background contribution is from the decay of $^{214}$Pb into the ground state of $^{214}$Bi, which occurs via a beta-decay with Q-value of 1.02\,MeV. The branching ratio (BR) for this channel, from the imported ENSDF, is 10.9\%. The shape of the beta spectrum at low energy reflects the calculation in ref.~\cite{Haselschwardt:2020iey}. We neglect the contribution from $^{214}$Bi beta-decay, as it can be identified by its short time separation from the subsequent $^{214}$Po alpha decay. In this study, we assume the XENONnT target $^{222}$Rn concentration of \SI{1}{\micro\becquerel/\kilogram}. This design goal is based on the \SI{4.5}{\micro\becquerel/\kilogram} level reached in XENON1T~\cite{xe1t-rnpaper}, ongoing radon emanation measurements of new detector components (xenon recirculation pumps, radon distillation column, TPC, inner cryostat, cables), and the estimated performance of the online radon distillation column. The expected background rate due to $^{222}$Rn is the largest contribution to the total rate in the energy ROI (see table~\ref{tab:bkg-rates-summary}) and amounts to 4.6\,\dru. We assign a 10\% systematic uncertainty to this prediction, driven by the uncertainty in the BR of the $^{214}$Pb beta-decay~\cite{WU2009681, ensdf}.

The radon isotope $^{220}$Rn can similarly emanate from materials as part of the $^{232}$Th decay chain. The beta-emitting $^{212}$Pb, product of this isotope, can contribute to background events at low energies. 
In XENON1T, we measured a $^{220}$Rn concentration relative to $^{222}$Rn of $\sim 0.3\%$. Assuming this relative concentration, the corresponding rate of $^{212}$Pb events is about 1\% of that expected from $^{214}$Pb.
We therefore neglect the $^{220}$Rn contribution in this work.

\subsubsection*{Krypton}
The xenon target contains natural krypton and therefore traces of the radioactive isotope $^{85}$Kr, which has a half-life of 10.76\,y. The decay of $^{85}$Kr via beta emission, with an end-point (kinetic) energy of 687\,keV, can contribute to the intrinsic background in the low-energy search region. The operation of the cryogenic distillation column in XENON1T reduced the \textsuperscript{nat}Kr concentration by 5 orders of magnitude~\cite{murra2018}. A minimum concentration of 0.36\,ppt\,(mol/mol), as measured with rare-gas mass spectrometry~\cite{Lindemann:2013kna}, was achieved during XENON1T's first science run~\cite{Aprile:2017iyp}. The measured relative abundance $^{85}$Kr/\textsuperscript{nat}Kr was $(1.7\pm 0.3)\times 10^{-11}$\,(mol/mol)~\cite{Aprile:2019dme}, determined using early high-krypton concentration XENON1T data. This is in agreement with earlier measurements from XENON100~\cite{Aprile:2017fhu}.We simulate the $^{85}$Kr decay rate in XENONnT based on this ratio and assuming the target concentration of 0.1\,ppt \textsuperscript{nat}Kr/Xe. This is within the reach of the distillation system which demonstrated a \textsuperscript{nat}Kr/Xe concentration level $<48$\,ppq~\cite{Aprile:2016xhi}. The entire xenon inventory of XENONnT was distilled through the krypton distillation column in 2019. We use the shape of the beta spectrum recently calculated in ref.~\cite{aprile2020lowER}. This results in a differential background rate of 1.1\,\dru~due to $^{85}$Kr, a factor 5 lower than $^{222}$Rn, with an uncertainty of 6\% on the spectral shape at low energy~\cite{Mougeot:2015bva}.

\subsubsection*{Xenon} 
Unstable xenon isotopes are distributed uniformly within the target volume. The long-lived $^{136}$Xe (t\textsubscript{1/2}\,=\,$2.17\times10^{21}$\,y~\cite{Albert:2013gpz}), with 8.9\% natural abundance, contributes to the background rate in the WIMP search region through double beta-emission (Q-value of 2.46\,MeV). We adopt the shape of the beta spectrum from ref.~\cite{KotilaIachelloWebSite}, which yields an average rate of 1.3\,\dru\,in the (1, 13)\,keV energy range. We assume an associated systematic uncertainty of 15\% due to the limited knowledge of the low-energy double beta spectrum~\cite{Kotila:2012zza}. The short-lived (t\textsubscript{1/2}\,=\,36\,d) $^{127}$Xe, which is produced by cosmic-ray activation, does not significantly contribute to the background.

The decay of $^{124}$Xe via double electron capture with emission of two neutrinos, first observed in XENON1T~\cite{XENON:2019dti}, gives rise to a new source of background. The detected signal is due to the cascade of X-rays and Auger electrons emitted as the vacancies in the lower electron shells are refilled from higher shells. In the rare case where both electrons are captured from the L-shell (BR\,=\,1.7\%), the expected energy deposition is $\sim$\,9.8\,keV. Two more electron capture lines at $\sim$\,36.7 keV (BR\,$\approx$\,23\%) and 64.3 keV (BR\,=\,75\%) are well outside the energy ROI. We assume the measured half-life of $1.4\times10^{22}$\,y and natural isotopic abundance of $0.095\%$. The resulting expectation rate for the LL-line in the energy ROI is 3.7\,\si{(t.y)^{-1}}. We assume a 30\% uncertainty on this rate, driven by the uncertainty in the measured half-life of $^{124}$Xe.

\begin{figure}[tbp]
    \centering
    \includegraphics[width=\textwidth]{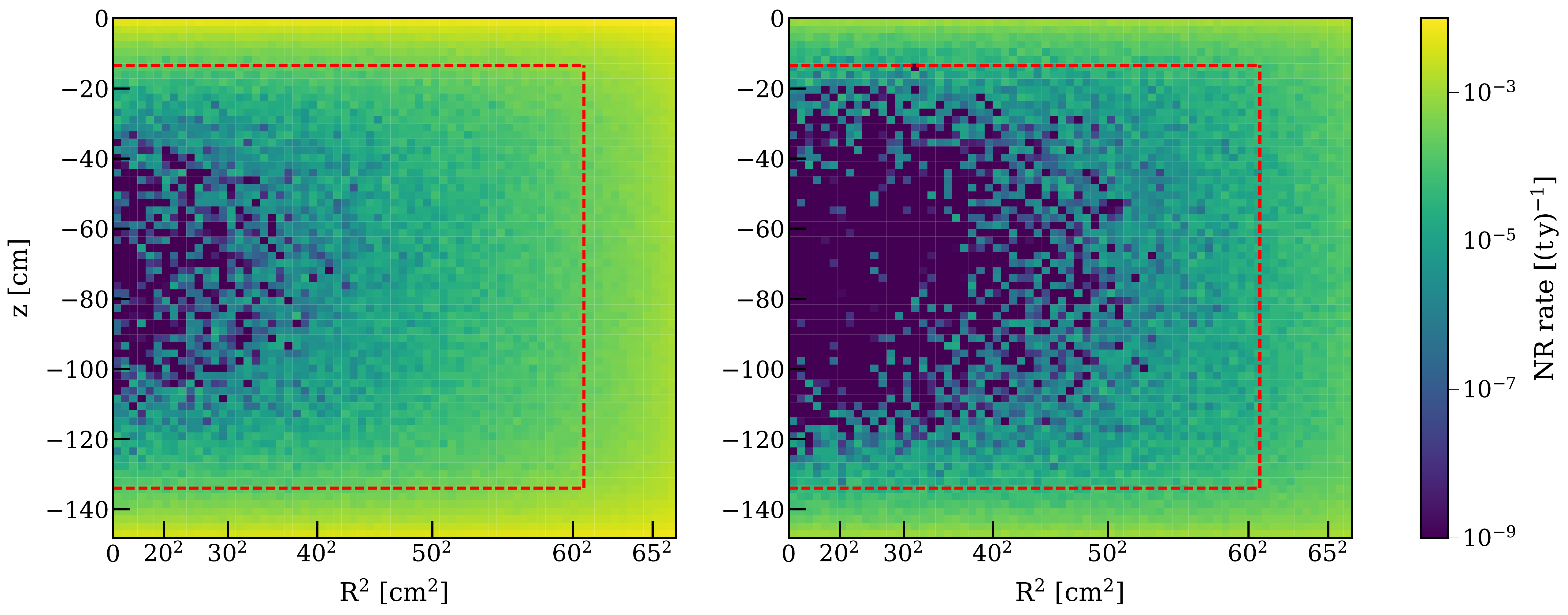}
    \caption{\label{fig:nr-dist} Spatial distribution of the radiogenic neutron background events inside the detector active region, in the (4,\,50)\,keV energy window, before (left) and after the NV cut is applied (right). The dashed red line corresponds to the 4\,t cylindrical fiducial volume. In this region, the total NR contribution from neutrons corresponds to 0.321\,\si{(t.y)^{-1}} without the suppression provided by the NV and 0.041\,\si{(t.y)^{-1}} after applying the NV cut.}
\end{figure}

\subsubsection*{Solar neutrinos} 
Solar neutrinos can elastically scatter off atomic electrons of the LXe target yielding low-energy ER signals. In contrast to ref.~\cite{Aprile:2015uzo}, where the free electron approximation was assumed, we take into account the effect of the atomic binding of electrons through the stepping approximation~\cite{Chen:2016eab} and conservatively assume a 3\% uncertainty, as suggested by the authors. The inclusion of this effect causes a suppression of the observable neutrino rate of about 20\% in the energy ROI. We consider the dominant sources of solar neutrinos: \emph{pp} fusion and electron capture by $^{7}$Be, which account for 98\% of the total neutrino flux. The estimated average contribution of 2.8\,\dru is the second largest source of ER background in XENONnT. Future multi-tonne scale LXe dark matter detectors can provide high-precision measurements of the low-energy solar neutrino flux~\cite{Baudis:2013qla, Aalbers:2020gsn}, contingent upon further reduction of the $^{222}$Rn background.

\subsection{Nuclear recoil background}\label{sec:nr-background}

\subsubsection*{Radiogenic neutrons} 
Radiogenic neutrons are produced through spontaneous fission (SF) or ($\alpha$, n) reactions in detector materials. Neutron yields and energies are calculated using the SOURCES-4A software~\cite{sources-4a, sources-4a-10MeV} as detailed in ref.~\cite{Aprile:2015uzo}, where the rates have been conservatively estimated by simulating the emission of a single neutron with no coincident gamma ray.  Events which produce a single elastic scatter in the active LXe volume are selected and weighted by the specific activities of the corresponding material and its neutron yield.

Due to the approximately 6\,cm-thick layer of LXe between the bottom PMT windows and the cathode, there is $\sim$\,250\,kg of LXe in which a neutron can scatter elastically producing an S1 signal without an associated S2. This region of the TPC is only sensitive to S1 light since there is no electric field to drift the ionization electrons towards the liquid-gas interface. If, in addition, the neutron scatters within the fiducial volume, the two prompt scintillation signals are observed as a single S1 signal due to the $\mathcal{O}$(ns) time of flight of MeV neutrons in LXe. These events, referred to as neutron-X, distort the neutron background distribution in the observable (cS1, cS2\textsubscript{b}) space, as they have lower S2/S1 ratios due to the additional S1 contribution from the S2-insensitive volume. Therefore, we also select events with a single elastic scatter in the fiducial volume and an energy deposition in the LXe below the cathode, where the combined energy falls within our NR energy ROI. Due to the much smaller mean-free path of gammas in LXe, the contribution from gamma-X events in our FV is negligible.

For the NV we account for signal smearing, wavelength-dependent QE corrections and the single photoelectron resolution in our NV PMT response model. For the results presented here, an NR background event is considered as tagged by the NV if at least 10 PMTs in coincidence record a signal above 0.5\,PE. In addition, we require that the NV signal occurs within a veto window of \SI{150}{\micro\second} from a single scatter observed in the TPC. The resulting NV tagging efficiency is approximately 87\%, corresponding to a reduction from 0.321\,\si{(t.y)^{-1}} to 0.041\,\si{(t.y)^{-1}} in the energy ROI and the 4 t FV. The gamma radiation due to the specific activity of all the materials surrounding the NV has also been assessed and was used to estimate the expected background induced in the NV. The overall background rate amounts to about 100\,Hz assuming the same tagging selection criteria, where the largest contribution is due to the radioactivity of the NV PMTs. In combination with the \SI{150}{\micro\second} veto window, this induces approximately 1.5\% of dead-time due to accidental coincidences between NV and TPC.

The spatial distribution of the radiogenic NR background in the TPC is shown in figure~\ref{fig:nr-dist}. The largest fraction of background events comes from the SS cryostat (27\% and 9\% of the total rate from the shells and flanges, respectively), the PMTs (33\% of the total, of which almost 60\% originate from the ceramic stem) and PTFE components (26\%). Copper components contribute less than 2\%, while the remaining 4\% is shared among the diving bell, SS electrode frames, NV ePTFE reflectors, and other components further away from the TPC. The estimated systematic uncertainty of the radiogenic neutron background rate is 50\%, accounting for the uncertainties on the neutron yields~\cite{sources-4a} and particle transportation models (comparing Geant4 and MCNP~\cite{Lemrani:2006dq} codes).

\subsubsection*{Cosmogenic neutrons}
Neutrons induced by cosmic muons interacting in the rock and concrete surrounding the detector can be tagged using the active muon veto in the water tank~\cite{Aprile:2014zvw}. 
The background of cosmogenic neutrons in the WIMP search region was suppressed to $<0.01$\,\si{(t.y)^{-1}} in XENON1T by tagging showers induced by external muon interactions~\cite{Aprile:2017iyp}. The addition of the Gd-loaded active NV will further reduce the rate of cosmogenic neutrons. Therefore, this source of background is not included in the XENONnT sensitivity estimation.

\begin{table}[t]
    \centering
    \begin{tabular}{lc}
    \toprule
    Source & Rate [\si{(t.y)^{-1}}]\\
    \midrule
    \multicolumn{2}{l}{\textbf{ER background}}\\
    Detector radioactivity & $\hphantom{0}25 \pm 3$\\
    $^{222}$Rn & $\hphantom{0}55 \pm 6$\\
    $^{85}$Kr & $\hphantom{0}13 \pm 1$\\
    $^{136}$Xe & $\hphantom{0}16\pm 2$\\
    $^{124}$Xe & $\hphantom{00}4 \pm 1$\\
    Solar neutrinos & $\hphantom{0}34 \pm 1$\\
    Total & $148 \pm 7$\\
    \midrule
    \multicolumn{2}{l}{\textbf{NR background}}\\
    Neutrons &  $(4.1 \pm 2.1) \times 10^{-2}$\\
    CE$\nu$NS (Solar $\nu$) &  $(6.3 \pm 0.3) \times 10^{-3}$\\
    CE$\nu$NS (Atm+DSN) &  $(5.4 \pm 1.1) \times 10^{-2}$\\
    Total & $(1.0 \pm 0.2) \times 10^{-1}$\\
    \bottomrule
    \end{tabular}
    \caption{\label{tab:bkg-rates-summary} Estimated background event rates in the 4\,t fiducial volume of XENONnT,  based on the energy of the recoil event. The energy ROI in which the event rates are integrated is (1,\,13)\,keV for ERs, and (4,\,50)\,keV for NRs. We assume  an activity concentration of \SI{1}{\micro\becquerel/\kilogram} of $^{222}$Rn and \SI{0.1}{ppt} (mol/mol) \textsuperscript{nat}Kr/Xe. The background contributions from Xe isotopes are determined assuming the 8.9\% and 0.095\% natural abundances of $^{136}$Xe and $^{124}$Xe, respectively.}
\end{table}

\subsubsection*{CE$\nu$NS neutrinos}
Neutrino interactions from solar, atmospheric and diffuse supernova (DSN) neutrinos contribute to the NR background through CE$\nu$NS. Solar neutrino backgrounds (predominantly $^{8}$B  and \textit{hep} neutrinos~\cite{Billard:2013qya}) limit the sensitivity to WIMPs with masses of a few GeV/c$^{2}$. In contrast, the NR spectra induced by atmospheric~\cite{Newstead:2020fie} and DSN~\cite{Billard:2013qya} neutrinos, shown in figure~\ref{fig:er-spectra} (right), extend to higher energy and affect the sensitivity to heavier WIMPs. Therefore, we distinguish two CE$\nu$NS components for the XENONnT background model: solar neutrinos ($^{8}$B+\emph{hep}) and the sum of atmospheric and diffuse supernova neutrinos. In the (4,\,50)\,keV energy range the expected rate of CE$\nu$NS from atmospheric (DSN) neutrinos is  4.8$\times$10$^{-2}$ (5.6$\times$10$^{-3}$)\,\si{(t.y)^{-1}}. The $^{8}$B spectrum is negligible above the 4\,keV lower bound, while \emph{hep} neutrinos induce a rate of only 6.3$\times$10$^{-3}$\,\si{(t.y)^{-1}}. However, Poisson fluctuations of the number of emitted scintillation photons can result in detection of events below the energy threshold. This is accounted for when we produce the complete background models in the observable (cS1, cS2\textsubscript{b}) space. Consequently, the solar neutrino (cS1, cS2\textsubscript{b}) distribution partially falls inside the WIMP search observable ROI and, given the much higher flux, $^8$B neutrinos become the dominant contribution to the total CE$\nu$NS background rate.

The systematic uncertainty associated with the solar neutrino CE$\nu$NS prediction is given by the uncertainty on the $^8$B neutrino flux (4\%)~\cite{Aharmim:2011vm}. We assume a 20\% uncertainty on the atmospheric and DSN neutrino rate, driven by the limited knowledge of the atmospheric neutrino flux~\cite{Honda:2011nf}.

\subsection{Summary}

The total expected background rate in the inner 4\,t FV is reported in table~\ref{tab:bkg-rates-summary}. In the energy ROI, the expected ER differential background rate is $12.3 \pm 0.6$\,\dru, a factor 6 lower than the $76\pm 2$\,\dru~measured in XENON1T~\cite{aprile2020lowER}. The NR differential background rate amounts to $(2.2\pm 0.5)\times 10^{-3}$\,\dru. 
The rates discussed in this section do not account for any ER discrimination. 
The complete detector response model, described in sections~\ref{sec:lxe-effects} and~\ref{sec:detector-effects}, is applied to the estimates described in this section to construct the background models in the observable (cS1, cS2\textsubscript{b}) space, used for the sensitivity projections. 
The final background expectations are discussed in section~\ref{sec:s1-s2-models} and summarized in table~\ref{tab:sensimus}.

\newcommand{\lagr}{\mathcal{L}} 
\newcommand{\Poiss}[2]{\mathrm{Pois}(#1 | #2)}
\newcommand{\xmeas}{\boldsymbol{x}}
\newcommand{\nuis}{\theta}
\newcommand{\nuiss}{\boldsymbol{\nuis}}
\newcommand{\Gauss}[3]{\mathrm{Gaus}(#1 | #2,#3 )}
\newcommand{\nuissh}{\hat{{\nuiss}}}
\newcommand{\nuisshah}{\hat{\nuissh}}

\section{Projected sensitivity}\label{sec:sensitivity}

We estimate the physics reach of XENONnT with the profile likelihood method~\cite{Cowan:2010js}. The statistical model is adapted from the procedure detailed for XENON1T, fully described in ref.~\cite{Aprile:2019dme}, with the exception of the spatial dimension which is not modeled in this work. The signal and background distributions are  defined in the two-dimensional (cS1, cS2\textsubscript{b}) space.
The target exposure is \SI{20}{\ton\year}, product of the assumed 4\,t fiducial mass and 5\,y livetime. With access to calibration data to characterize the background population close to the TPC edges, as done for the XENON1T analysis, XENONnT may be able to extend the fiducial mass beyond the reference 4\,t cylinder.

To simulate the detection and data selection efficiencies, we apply the combined XENON1T efficiency curve~\cite{Aprile:2018dbl,Aprile:2019bbb}, including a three-fold S1 coincidence requirement, and a cS1 range of (3,\,100)\,PE, to the background spectra presented in section~\ref{sec:backgrounds}. This corresponds to an average acceptance of 82\% (83\%) for ER (NR) events in the chosen cS1 range. The cS1 range corresponds to the energy ranges used in section~\ref{sec:backgrounds} and defines our region of interest in the observable (cS1, cS2\textsubscript{b}) space (observable ROI). 

\subsection{Background and WIMP signal models}\label{sec:s1-s2-models}

The background of XENONnT is modeled as four components: total ER (materials, $^{222}$Rn, $^{85}$Kr, solar neutrinos, $^{136}$Xe and $^{124}$Xe), radiogenic neutrons and  CE$\nu$NS, split into solar ($^8$B+\emph{hep}) and the sum of atmospheric and diffuse supernova neutrinos. The respective recoil energy spectra, discussed in section~\ref{sec:backgrounds}, are converted into (cS1, cS2\textsubscript{b}) distributions by applying the detector response model described in sections~\ref{sec:lxe-effects} and \ref{sec:detector-effects}.

\begin{figure}[t]
    \centering
    \includegraphics[width=\textwidth]{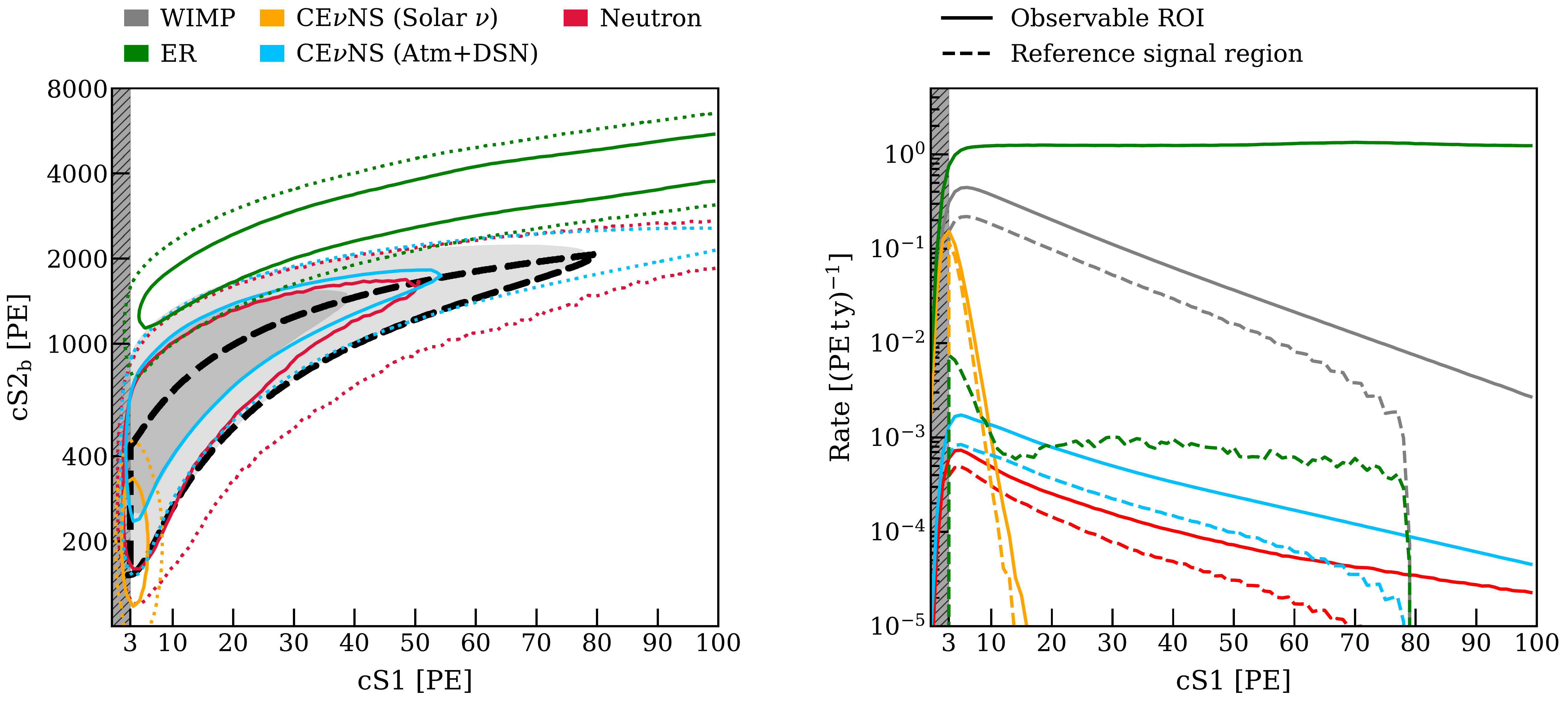}
    \caption{\label{fig:final_models}
    (Left) Background and WIMP signal PDFs in the (cS1, cS2\textsubscript{b}) space. The 1$\sigma$ (solid) and 2$\sigma$ (dotted) contours are shown for the background components. The distribution of a 50\,GeV/c$^2$ WIMP is drawn as a dark gray (1$\sigma$) and gray (2$\sigma$) filled region. The black dashed contour line defines the reference signal region, bounded by the 2$\sigma$ closed WIMP region below the median cS2\textsubscript{b}. The shaded region is outside the (3,\,100)\,PE observable ROI and is excluded from sensitivity estimates. (Right) Background and signal distributions projected onto the cS1 space (solid curves). Dashed lines indicate the reduced rate of each component inside the reference signal region highlighted in the left plot. The shape of the ER spectrum inside the reference signal region is driven by the cS1-dependent discrimination power. The 50\,GeV/c$^2$ WIMP rate assumes a cross-section of $\sigma_\mathrm{DM}$\,=\,\SI{5e-47}{\cm\squared}.}
\end{figure}

The WIMP signal NR spectrum is derived assuming a standard isothermal dark matter halo with density ~$\rho_{\mathrm{DM}}$\,=\,0.3\,GeV/c$^3$, most probable WIMP velocity $v_0$\,=\,220\,km/s, escape velocity ~$v_{\mathrm{esc}}$\,=\,544\,km/s and uses the Helm form factor for the nuclear cross-section~\cite{LEWIN199687}. Spectra are produced for WIMP masses between 6\,GeV/c$^2$ and 10\,TeV/c$^2$. WIMP interactions are assumed to be uniformly spread throughout the TPC. The probability density functions (PDFs) of the four background models and a 50\,GeV/c$^2$ WIMP signal are shown in figure~\ref{fig:final_models} (left) along with their projections onto cS1 (right).

\begin{table}[t]
    \centering
    \begin{tabularx}{\linewidth}{lccc}
    \toprule
    Model component & \multicolumn{2}{c}{Expectation value ($\mu$) in \SI{20}{\ton\year}} & \multicolumn{1}{p{2.8cm}}{\centering Rate uncertainty} \\ \cline{2-3}
    & \parboxc{c}{3ex}{Observable ROI} & \parboxc{c}{3ex}{Reference signal region} & ($\xi$)\\
    \midrule
    \textbf{Background}\\
    ER & 2440 & 1.56 &\\
    Neutrons & 0.29 & 0.15 & 50\%\\
    CE$\nu$NS (Solar $\nu$) & 7.61 & 5.41 & \hphantom{0}4\%\\
    CE$\nu$NS (Atm+DSN) & 0.82 & 0.36 & 20\%\\
    \midrule
    \textbf{WIMP signal}\\
    6\,GeV/c$^2$ \kern 5pt ($\sigma_\mathrm{DM}=3\times10^{-44 }\,\mathrm{cm}^2)$ &\hphantom{0}25 & \hphantom{0}19 &\\
    50\,GeV/c$^2$ ($\sigma_\mathrm{DM}=5\times10^{-47 }\,\mathrm{cm}^2)$ & 186 & \hphantom{0}88 &\\
    1\,TeV/c$^2$ \kern 7pt ($\sigma_\mathrm{DM}=8\times10^{-46 }\,\mathrm{cm}^2)$ & 286 & 118 &\\
    \bottomrule
    \end{tabularx}
    \caption{\label{tab:sensimus}Expected number of events in the (3,\,100)\,PE cS1 observable ROI, for the \SI{20}{\ton\year} target exposure of XENONnT. The rates take into account signal fluctuation. These Poisson fluctuations are of particular importance to the CE$\nu$NS (Solar $\nu$) rate, and result in detection of events below the nominal energy threshold used in table~\ref{tab:bkg-rates-summary}. Detection and selection efficiencies are also accounted for. We show results for the background components included in the statistical model as well as for 6, 50 and 1000\,GeV/c$^2$ WIMP signals. The cross-sections are chosen to be close to the XENON1T exclusion limit~\cite{Aprile:2018dbl}. Expectation values in the reference signal region reflect the residual fraction of each model component falling inside the $2\sigma$ contour of the 50\,GeV/c$^2$ WIMP PDF, below the cS2\textsubscript{b} median. Background uncertainties, where the rate is constrained by ancillary measurement terms included in the full likelihood, are reported in the last column. The ER rate will be highly constrained by data, thus no uncertainty is included.} 
\end{table}

In XENON1T 99.7\% ER discrimination~\cite{Aprile:2018dbl} was achieved in a reference region below the median of a 200\,GeV/c$^2$ WIMP (signal) PDF in the observable ROI. Assuming the same reference region in XENONnT, the resulting ER rejection from our emission and detector model is 99.9\%. The increase in the projected rejection power of XENONnT is mainly driven by the expected increase in electron lifetime due to the novel LXe purification system and the higher drift field of 200\,V/cm. Under reasonable variations of those parameters, and consequently the ER rejection level, the changes in the expected sensitivity are within $\sim$\,10\% for a 50\,GeV/c$^2$ mass WIMP. Uncertainties from the final detector conditions are thus subdominant to the statistical uncertainty due to the low event rate regime of WIMP searches.

Expectation values of the signal and background components for a 20\,\si{t.y} exposure inside the 4\,t FV are listed in table~\ref{tab:sensimus}. This table also includes expectation values in the reference WIMP signal region, indicated by a dashed grey line in figure~\ref{fig:final_models} (left). This reference signal region is defined as the area below the median cS2\textsubscript{b} of a 50\,GeV/c$^2$ spin-independent WIMP signal, and within the 2$\sigma$ contour. The expected fraction of ER events falling inside this region corresponds to $6\times10^{-4}$. The neutron background distribution overlaps with the reference signal region by 54\%, slightly more than atmospheric neutrinos and DSN (44\%), due to the impact of the neutron-X population. A fraction of 71\% of the CE$\nu$NS PDF from solar neutrinos falls inside the reference signal region, even though it is confined to very small cS1 and cS2\textsubscript{b} signals. Numbers in this portion of the observable space can only give an indication of performance, but are useful for comparison with other detectors. The sensitivity study presented below does not use any ER discrimination cut or specific signal region selection, but it is based on the profile likelihood analysis in the full (cS1,  cS2\textsubscript{b}) observable ROI. 

The neutron and CE$\nu$NS background rates are primarily constrained by ancillary measurements, as discussed in section~\ref{sec:nr-background}, and likelihood terms are included to account for the relative uncertainties reported in table~\ref{tab:sensimus}. On the other hand, even a short first run of XENONnT will constrain the ER rate better than the 10\% prediction uncertainty, therefore we do not include a related term in the likelihood. Systematic uncertainties on the detector response to NRs primarily impact the search for low-mass WIMPs. However, such uncertainties were sub-dominant in the XENON1T WIMP search results and we therefore neglect them in this work.

\subsection{Statistical model}\label{sec:statistical-model}

\begin{figure}[t]
    \centering
    \includegraphics[width=\textwidth]{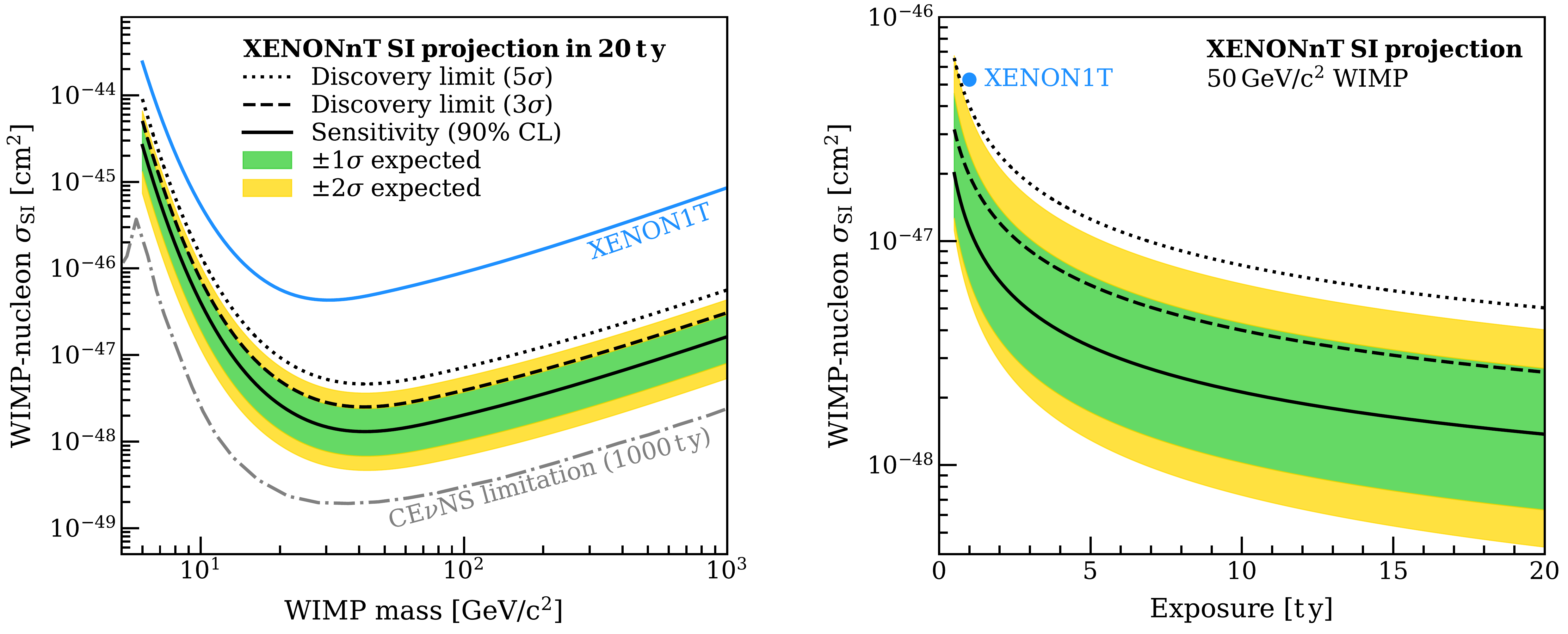}
    \caption{\label{fig:sensitivity_projections} Projections of the XENONnT sensitivity and discovery power in the search for spin-independent WIMP-nucleon couplings. (Left) Median $90\%$ CL exclusion limit (black solid line) for a \SI{20}{\ton\year} exposure, with the $1\sigma$ (green) and $2\sigma$ (yellow) bands. The current strongest exclusion limit, obtained with XENON1T~\cite{Aprile:2018dbl}, is shown in blue. The gray dashed-dotted line represents the discovery limit of an idealized LXe-based experiment with CE$\nu$NS as unique background source and a \SI{1000}{t.y} exposure~\cite{Ruppin_2014}. The improvement of the discovery potential with increasing exposure below that line would be significantly slowed down by the atmospheric neutrino background. (Right) Sensitivity as a function of exposure, for the search of a 50\,GeV/c$^2$ WIMP in the assumed \SI{4}{\tonne} fiducial mass. The dashed (dotted) black lines in both panels indicate the smallest cross-sections at which the experiment would have a $50\%$ chance of observing an excess with significance greater than $3\sigma$ ($5\sigma$). A two-sided profile construction is used to compute the confidence intervals.}
\end{figure}

The likelihood-based statistical modeling of the experiment uses an extended unbinned likelihood, $\lagr$, with PDFs in $\xmeas=\,$(cS1, cS2\textsubscript{b}): 
    \begin{align}\large
        \lagr(\sigma_\mathrm{DM},\nuiss) =\, & \Poiss{N}{\mu_\mathrm{tot}(\sigma_\mathrm{DM},\nuiss)} \cdot\prod_{i=1}^{N}\left[\sum_c\frac{\mu_c(\sigma_\mathrm{DM},\nuiss)}{\mu_\mathrm{tot}(\sigma_\mathrm{DM},\nuiss)}\cdot f_c(\xmeas_i|\nuiss)  \right]\cdot \lagr_\mathrm{anc}(\nuiss) \ ,
    \label{eqn:unbinned}
    \end{align}
   
\noindent where $\mu_\mathrm{tot}(\sigma,\nuiss) \equiv \,\, \sum_c \mu_c(\sigma,\nuiss)$ and the ancillary term $\lagr_\mathrm{anc}$ is defined as

    \begin{align}\large
        \lagr_\mathrm{anc}(\nuiss) \equiv \,&\,
        \prod_{k}\Gauss{\hat{\mu}_{k}}{\mu_{k}}{\xi_{k}} \ ,
    \label{eqn:unbinned_2}
    \end{align}
with $k$ running over the three background components with associated uncertainties, namely neutrons and CE$\nu$NS from solar and Atm+DSN neutrinos.

The likelihood, evaluated for each WIMP mass $M_\mathrm{DM}$, is a function of the WIMP cross-section $\sigma_\mathrm{DM}$ and nuisance parameters $\nuiss$, which parameterise the PDFs $f_c$ and expectation values $\mu_c$. The index~$c$ runs over the background components and WIMP signal.
The rate uncertainties $\xi$ from ancillary measurements are taken into account as Gaussian constraints in the $\lagr_\mathrm{anc}$ term.
The observed events, indexed by $i$, are collected in a vector with length $N$. The profiled log-likelihood ratio for each considered WIMP mass
\begin{equation}
   q(\sigma_\mathrm{DM}) \equiv -2\cdot\log\frac{\lagr(\sigma_\mathrm{DM},\nuisshah)}{\lagr(\hat{\sigma}_\mathrm{DM},\nuissh)}
   \label{eqn:stat1}
\end{equation}
is used as a test statistic to test both the signal and null, $q(\sigma_\mathrm{DM}=0)$, hypotheses. The likelihood is maximised at $(\hat{\sigma}_\mathrm{DM},\nuissh)$, and $\nuisshah$ are the nuisance parameters that maximize the likelihood for a given $\sigma_\mathrm{DM}$. The distributions of $q(\sigma_\mathrm{DM})$ are estimated with  $\mathcal{O}(10^4)$ toy MC simulations of the experimental data, including both the science data and ancillary measurements.

The signal and background-only hypotheses testing (for the sensitivity and discovery limits, respectively) and the construction of confidence intervals follow the approach detailed in ref.~\cite{Aprile:2019dme}.
The adopted two-sided profile construction~\cite{Feldman:1997qc,PDG} of confidence intervals ensures correct coverage when switching from reporting one-sided (exclusion limits) to two-sided intervals (discovery). This is different from previous sensitivity projections based on a one-sided Neyman profile construction, which results in a systematically stronger sensitivity as discussed in appendix~\ref{sec:appendix_onesided}.

\begin{figure}[t]
    \centering
    \includegraphics[width=\textwidth]{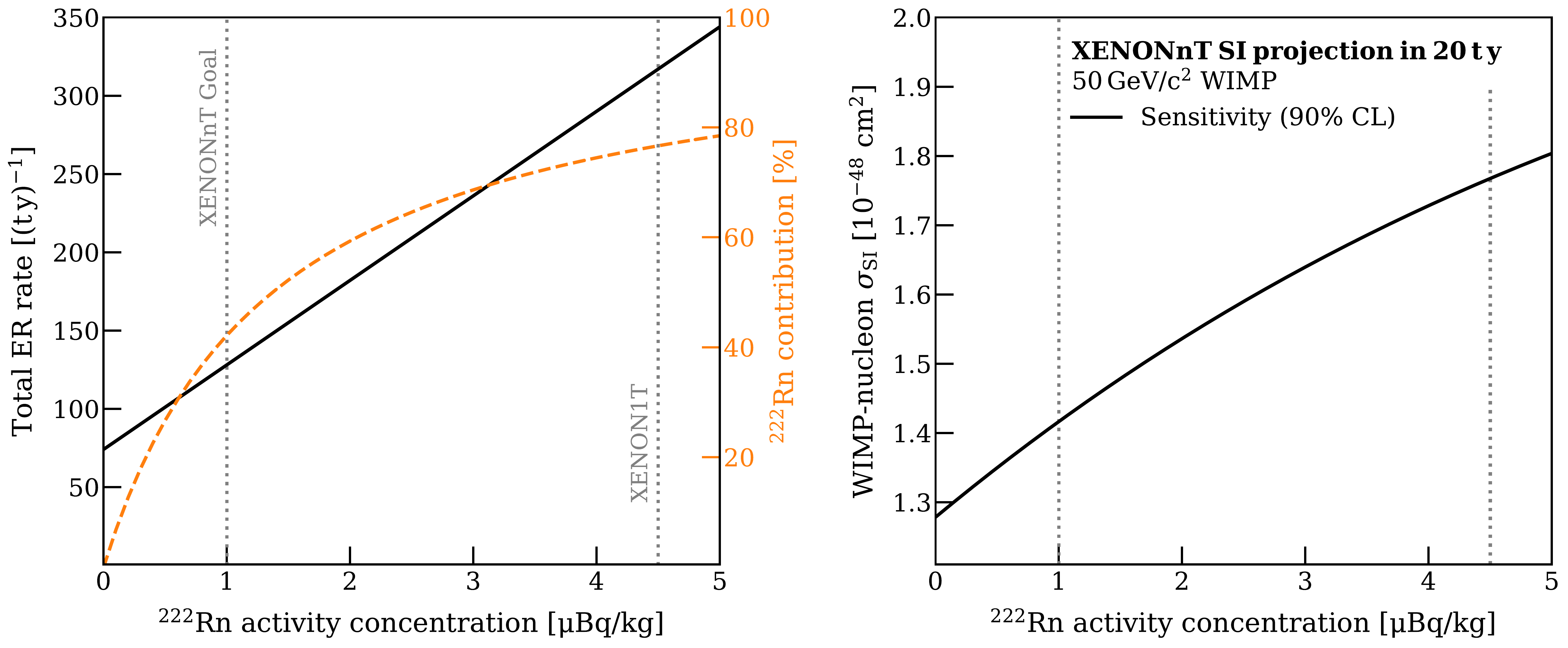}
    \caption{\label{fig:radon_sensitivity}
    (Left) ER background rate in the (3,\,100) cS1  observable ROI as a function of the $^{222}$Rn concentration in LXe. The orange dashed line represents the $^{222}$Rn fractional contribution to the total ER background. The dotted grey lines indicates the XENONnT goal of \SI{1}{\micro\becquerel/\kilogram} and the \SI{4.5}{\micro\becquerel/\kilogram} $^{222}$Rn concentration achieved in XENON1T~\cite{xe1t-rnpaper}. (Right) Projection of the XENONnT sensitivity to spin-independent couplings of a 50\,GeV/c$^2$ WIMP under varying assumptions on the $^{222}$Rn concentration.}
\end{figure}

\subsection{Sensitivity and discovery power}

The sensitivity presented in figure~\ref{fig:sensitivity_projections} (left) expresses the median exclusion limit at 90\% CL on the SI WIMP-nucleon cross-section. With its ultimate \SI{20}{\ton\year} exposure, XENONnT can probe cross-sections more than an order of magnitude below the current best limits set by XENON1T~\cite{Aprile:2018dbl}, reaching the strongest sensitivity of $1.4\times10^{-48}\,\mathrm{cm}^2$ for a 50\,GeV/c$^2$ WIMP. The projected XENONnT median discovery levels with $3\sigma$ (dashed) and $5\sigma$ (dotted) significance are shown along with the sensitivity (solid). The minimum WIMP cross-section at which the experiment has a 50\% chance of observing an excess with a significance greater than $3\sigma$ ($5\sigma$) is $2.6\times10^{-48}\,\mathrm{cm}^2$ ($5.0\times10^{-48}\,\mathrm{cm}^2$), corresponding to a mass of 50\,GeV/c$^2$. In figure~\ref{fig:sensitivity_projections} (right), we also report the sensitivity and discovery power for a 50\,GeV/c$^2$ WIMP as a function of exposure.

The largest source of ER background events arises from $^{222}$Rn. In figure~\ref{fig:radon_sensitivity}, we evaluate the XENONnT sensitivity to a 50\,GeV/c$^2$ WIMP for $^{222}$Rn activity concentrations ranging up to \SI{5}{\micro\becquerel/\kilogram}. The XENONnT goal of \SI{1}{\micro\becquerel/\kilogram} and the \SI{4.5}{\micro\becquerel/\kilogram} $^{222}$Rn activity concentration achieved with XENON1T~\cite{xe1t-rnpaper} are indicated by grey dotted lines. At \SI{4.5}{\micro\becquerel/\kilogram} the sensitivity is $\sim$25\% worse than at the XENONnT goal of \SI{1}{\micro\becquerel/\kilogram}.

\begin{figure}[t]
    \centering
    \includegraphics[width=\textwidth]{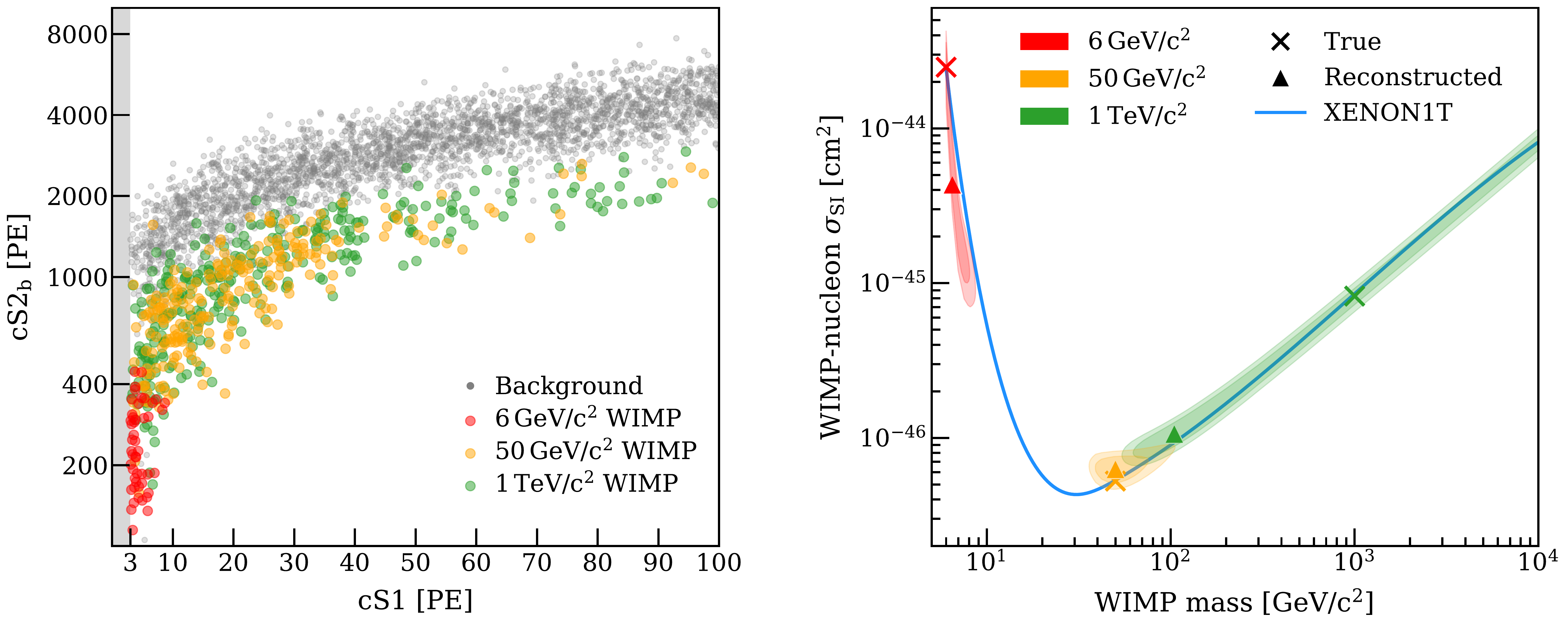}
    \caption{\label{fig:discoveryplots} Signal events for WIMPs with masses of 6\,GeV/c$^2$ (red), 50\,GeV/c$^2$ (orange) and 1\,TeV/c$^2$ (green) in a \SI{20}{t.y} exposure. The considered best-case scenario spin-independent cross-sections, equal to those used in table~\ref{tab:sensimus}, correspond to values close to the XENON1T upper limit (blue line). (Left) Distribution of signal events in the observable (cS1, cS2\textsubscript{b}) space, with background events indicated as gray circles. (Right) $1\sigma$ and $2\sigma$ confidence contours for each excess, with a triangle for the best-fit point and a cross marker indicating the true value used to generate each dataset.}
\end{figure}

To illustrate the precision with which  $\sigma_\mathrm{DM}$ and $M_\mathrm{DM}$ could be reconstructed in case of a discovery, we generate three toy signal datasets, shown in figure~\ref{fig:discoveryplots} (left), for excesses generated by 6\,GeV/c$^2$, 50\,GeV/c$^2$, and 1\,TeV/c$^2$ WIMPs, with cross-sections chosen as in table~\ref{tab:sensimus}, close to the XENON1T upper limits, and a \SI{20}{\ton\year} exposure. Contours in ($\sigma_\mathrm{DM}, M_\mathrm{DM}$) are computed with the asymptotic assumption that $q(\sigma_\mathrm{DM},M_\mathrm{DM})$ is distributed according to a $\chi^2$-distribution with two degrees of freedom. Figure~\ref{fig:discoveryplots} (right) shows the $1\sigma$ and $2\sigma$ constrained regions for the three excesses. For low-mass WIMPs, constraints on the cross-sections can span more than one order of magnitude, while the mass reconstruction precision is high. For WIMP masses around the projected sensitivity minimum well-constrained two-sided intervals can be obtained. With increasing masses, the WIMP spectra become degenerate and inference results may be scaled according to the WIMP mass, resulting in unconstrained contours at high WIMP masses and cross-sections.

\begin{figure}[t]
    \centering
    \includegraphics[width=\textwidth]{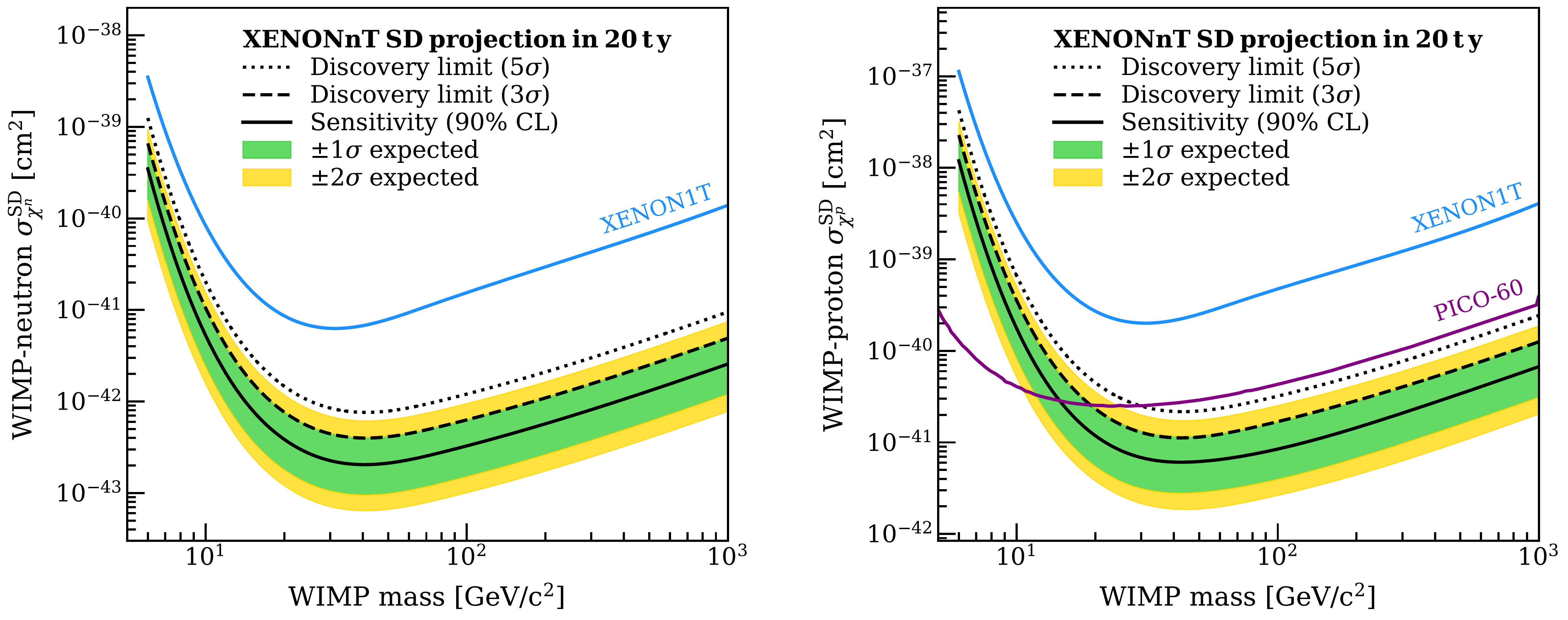}
    \caption{\label{fig:sensitivity_sd} Projections of the sensitivity of XENONnT to spin-dependent WIMP couplings to neutrons (left) and protons (right). Median $90\%$ CL exclusion limit (black solid line) for a \SI{20}{\ton\year} exposure, with the $1\sigma$ (green) and $2\sigma$ (yellow) bands. The dashed and dotted black lines represent the $3\sigma$ and $5\sigma$ discovery limits, respectively. The blue lines indicate the XENON1T upper limits~\cite{Aprile:2019dbj} and the solid purple line is the PICO-60 upper limit~\cite{picototal}.}
\end{figure}

In addition to coherently enhanced spin-independent scattering, spin-dependent interactions~\cite{LEWIN199687} are included in any non-relativistic theory of WIMP-nucleus scattering. Searches for this interaction are commonly constrained to the proton- and neutron-only cases. In figures~\ref{fig:sensitivity_sd}, we show the sensitivity of XENONnT to these interactions, using the same background models as in the spin-independent case, but utilising the signal recoil models of ref.~\cite{Aprile:2019dbj}. For a 20\,\si{t.y} exposure, the projected WIMP sensitivity of XENONnT to neutron (proton) couplings is $2.2\times10^{-43}\,\mathrm{cm}^2$ ($6.0\times10^{-42}\,\mathrm{cm}^2$) for a 50\,GeV/c$^2$ WIMP.

\section{Conclusions}\label{sec:conclusions}

The XENONnT direct WIMP detection experiment will start taking data in 2020. The newly developed neutron veto, LXe purification system and radon distillation column will further suppress backgrounds in the detector. A full model of the detector has been developed in Geant4 to estimate the physics reach of the experiment.

We estimated the ER and NR backgrounds based on the results of the material radioassay campaign, and using well-motivated assumptions for the xenon-intrinsic contaminants, such as $^{222}$Rn and $^{85}$Kr. We also studied the ability of the neutron veto detector to tag potential radiogenic neutron background events.  Prior to accounting for detector effects, selection efficiencies and signal fluctuations we predict a background rate of $12.3 \pm 0.6$\,\dru~and $(2.2\pm 0.5)\times 10^{-3}$\,\dru, for electronic and nuclear recoil events respectively. The full background and WIMP signal models were produced by converting the predicted recoil energy spectra into distributions in the observable cS1 and cS2\textsubscript{b} signal space. The adopted detector response model is based on the XENON1T LXe emission model, while detector-dependent parameters, such as LCE maps, electron lifetime and drift field intensity, are chosen for the XENONnT case. The sensitivity study was performed in the (cS1, cS2\textsubscript{b}) signal space,
which includes the entire ER distribution before S2/S1-based rejection. Taking into account the detector response and data selection efficiencies the expected ER (NR) background rate amounts to 122 (0.44)\,\si{(t.y)^{-1}}, which is reduced to 0.08 (0.30)\,\si{(t.y)^{-1}} in the reference signal region.

The XENONnT sensitivity to WIMP-nucleus interactions is projected using the profile likelihood ratio method, using a statistical model similar to that of the XENON1T data analysis~\cite{Aprile:2019dme}. A five-year search using a central 4\,t fiducial volume will push the sensitivity of the detector to spin-independent interactions to $1.4\times10^{-48}\,\mathrm{cm}^2$ for a 50\,GeV/c$^2$ WIMP, more than one order of magnitude beyond the current best limits, set by XENON1T. With the same \SI{20}{t.y} exposure, a 50\,GeV/c$^2$ WIMP with cross-section of $2.6\times10^{-48}\,\mathrm{cm}^2$ ($5.0\times10^{-48}\,\mathrm{cm}^2$) will yield a median $3\sigma$ ($5\sigma$) discovery significance. Similar improvements in sensitivity will also be achieved by XENONnT in the search for spin-dependent WIMP interactions. The unprecedented sensitivity of the XENONnT dark matter detector will allow us to probe large fractions of the yet unexplored regions of WIMP parameter space.
\appendix
\section{One-sided upper limit construction and sensitivity comparison}\label{sec:appendix_onesided}

In the XENONnT sensitivity study presented in this work, we adopted the same unified Feldman-Cousins confidence interval construction~\cite{Feldman:1997qc} used for the science results of XENON1T~\cite{Aprile:2019dme,Aprile:2018dbl}. This two-sided construction provides the correct coverage while yielding either one- or two-sided intervals depending on the experimental outcome. In contrast, using the one-sided construction below a chosen discovery significance threshold, and switching to two-sided interval construction above that, may result in undercoverage of the reported limit. This issue is known as the ``flip-flop'' problem.

\begin{figure}[t]
    \centering 
    \includegraphics[width=0.45\textwidth]{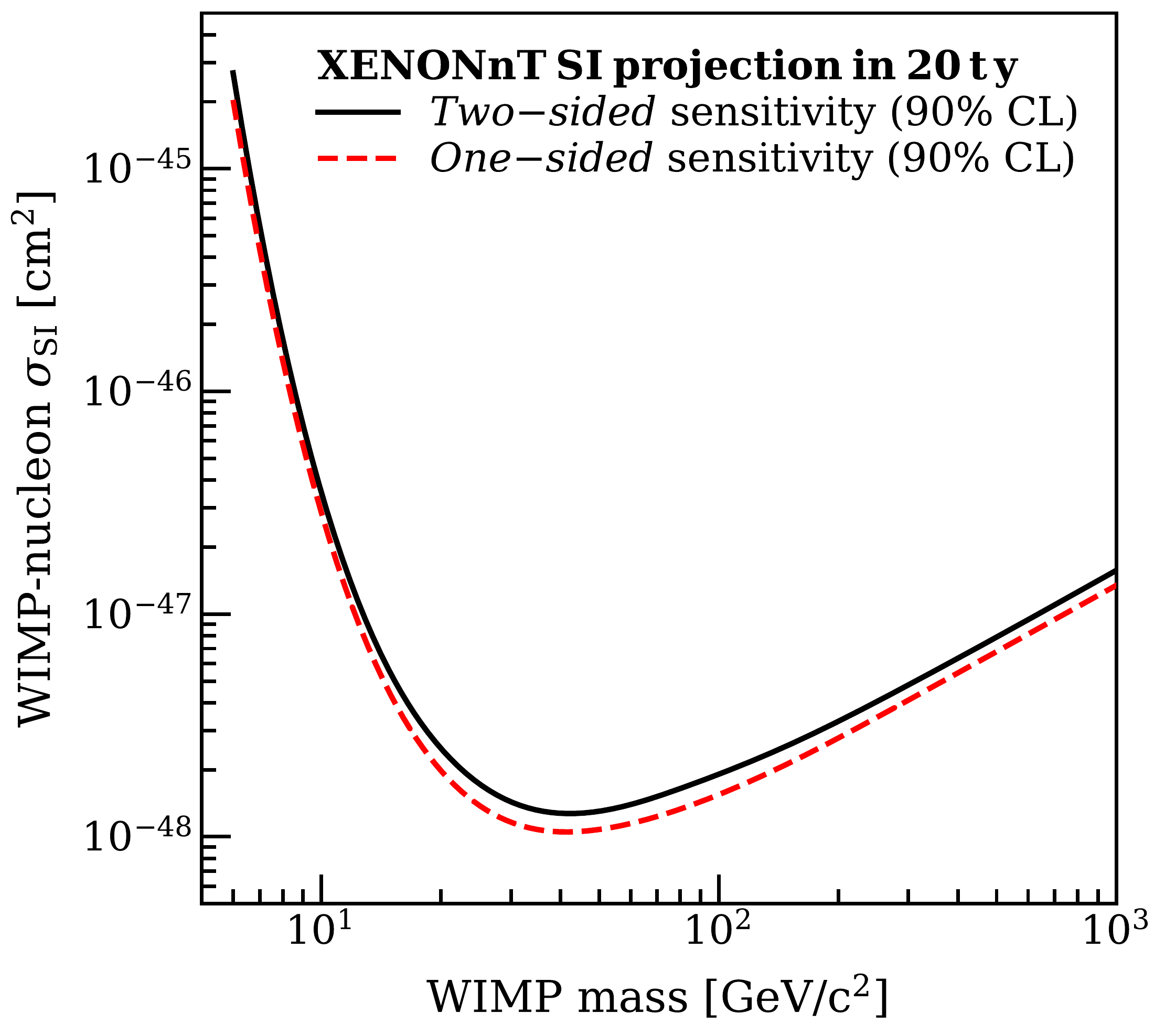}
    \caption{\label{fig:sensitivity_1sided} Comparison of sensitivity projections to spin-independent WIMP-nucleon couplings using one-sided upper limit construction (dashed red) and two-sided interval construction (black, taken from figure~\ref{fig:sensitivity_projections} left).}
\end{figure}

Projected sensitivities of direct-detection dark matter experiments are often reported using one-sided limit constructions~\cite{lz_sensi, Aprile:2015uzo}. The one-sided 90\% confidence level sensitivity for XENONnT for a \SI{20}{t.y} exposure is shown in figure~\ref{fig:sensitivity_1sided} and compared to the result in figure~\ref{fig:sensitivity_projections}. The difference between the one-sided and two-sided approaches can be considerable, up to $\sim$\,30\% across most of the considered mass range. Therefore, care should be taken when comparing the spin-independent and spin-dependent WIMP sensitivities presented in this work, produced with the two-sided confidence interval construction, with those of other works. As the choice of computing one-sided or unified confidence intervals must be made before unblinding to prevent bias, figure~\ref{fig:sensitivity_1sided} should not be used to scale experimental results.

\section{Impact of potential $^3$H background on XENONnT sensitivity}\label{sec:appendix_tritium}

Detailed studies on the low-energy ER background of XENON1T have shown a statistically significant excess in the region <\,7\,keV~\cite{aprile2020lowER}. Several possible origins for the observed excess were explored: new physics (such as solar axions or a neutrino magnetic moment), or the hypothesis that the excess is due to traces of $^3$H. 

\begin{figure}[t]
    \centering
    \includegraphics[width=\textwidth]{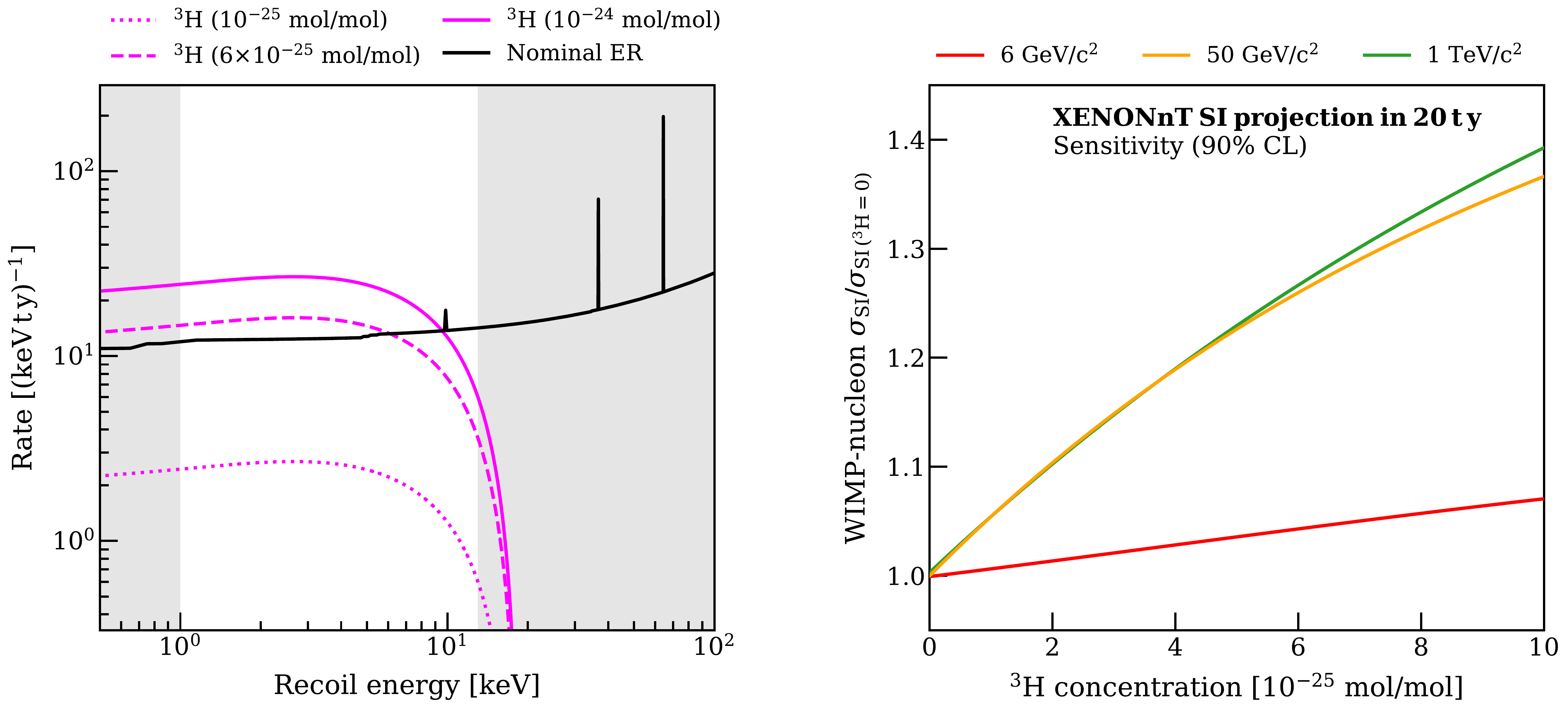}
    \caption{\label{fig:tritium_bkg} (Left) ER background energy spectra in the XENONnT 4\,t fiducial volume. The $^3$H background produced by a concentration of $10^{-24}$ (6$\times10^{-25}$, $10^{-25}$)\,mol/mol is shown by the magenta solid (dashed, dotted) line. The nominal ER background from all other sources is represented by the solid black line, taken from figure~\ref{fig:er-spectra}. (Right) Sensitivity for spin-independent WIMP-nucleon coupling in \SI{20}{t.y}, for WIMPs with masses of 6\,GeV/c$^2$ (red), 50\,GeV/c$^2$ (orange) and 1\,TeV/c$^2$ (green), as a function of the considered $^3$H concentrations. Projections are reported relative to the median sensitivity in fig.~\ref{fig:sensitivity_projections} (left), in which  we assume no $^3$H contribution.}
\end{figure}

We study here the impact of the Standard Model hypothesis. Since the exact dynamics of $^3$H within the LXe inventory, as it is recirculated, purified and cooled, are not fully understood, it is very difficult to make a prediction of the $^3$H content. We therefore evaluate the impact of possible $^3$H concentrations ranging from $10^{-24}$\,mol/mol to $10^{-25}$\,mol/mol, consistent with the assumption that the entire excess observed in XENON1T can be attributed to $^3$H. The intrinsic background induced in the 4\,t FV for the maximum assumed $^3$H contribution is shown in figure~\ref{fig:tritium_bkg} (left) as the solid magenta line. The best fit value for $^3$H in XENON1T ($6\times10^{-25}$\,mol/mol) and a $10^{-25}$\,mol/mol contribution is shown in the dashed and dotted magenta lines respectively. Concentrations in excess of $7\times10^{-25}$\,mol/mol would result in this component being larger than the nominal ER backgrounds in the energy ROI.

To assess the impact on the projected sensitivity of the experiment, we follow the procedure detailed in section~\ref{sec:sensitivity}, where the $^3$H is added as a fifth background component. In the \SI{20}{\ton\year} exposure, the highest tritium contribution would yield $\sim$\,3550 events in the observable ROI and 3.4 events within the reference signal region, twice that of all other ER backgrounds combined. The lowest considered $^3$H concentration would only increase the overall ER background by $\sim$\,20\%.

The impact on the sensitivity of XENONnT to spin-independent interactions is shown in figure~\ref{fig:tritium_bkg} (right) for three different WIMP masses. In order to disentangle the $^3$H  contribution from the nominal ER background in our observable ROI using XENONnT data, the nominal ER rate can be constrained from measurements above the end point of the $^3$H beta-spectrum (18\,keV). Therefore, in contrast to the procedure detailed in section~\ref{sec:sensitivity}, we add a Gaussian ancillary term to the likelihood function assuming a 10\% uncertainty on the nominal ER background. The $^3$H expectation value is left unconstrained. The impact on both the projected sensitivity and $3\sigma$ discovery potential is minimal for a 6\,GeV/c$^2$ WIMP; $\sim$\,10\% worse at the highest $^3$H concentration. 
An increased $^3$H background level has a larger impact on the sensitivity to higher WIMP masses. For 50\,GeV/c$^2$ and 1\,TeV/c$^2$ WIMPs, the sensitivity would be $\sim$\,40\% worse and the $3\sigma$ discovery limit would increase by $\sim$\,35\%.

\section*{Acknowledgements}
We gratefully acknowledge support from the National Science Foundation, Swiss National Science Foundation, German Ministry for Education and Research, Max Planck Gesellschaft, Deutsche Forschungsgemeinschaft, Netherlands Organisation for Scientific Research (NWO), Weizmann Institute of Science, ISF, Fundacao para a Ciencia e a Tecnologia, Région des Pays de la Loire, Knut and Alice Wallenberg Foundation, Kavli Foundation, JSPS Kakenhi in Japan and Istituto Nazionale di Fisica Nucleare. This project has received funding or support from the European Union’s Horizon 2020 research and innovation programme under the Marie Sklodowska-Curie Grant Agreements No. 690575 and No. 674896, respectively. Data processing is performed using infrastructures from the Open Science Grid, the European Grid Initiative and the Dutch national e-infrastructure with the support of SURF Cooperative. We are grateful to Laboratori Nazionali del Gran Sasso for hosting and supporting the XENON project.

\bibliography{references.bib}
\end{document}